\documentclass[pdflatex,sn-mathphys-num]{sn-jnl}

\usepackage{graphicx}%
\usepackage{multirow}%
\usepackage{amsmath,amssymb,amsfonts}%
\usepackage{amsthm}%
\usepackage{mathrsfs}%
\usepackage[title]{appendix}%
\usepackage{xcolor}%
\usepackage{textcomp}%
\usepackage{manyfoot}%
\usepackage{booktabs}%
\usepackage{algorithm}%
\usepackage{algorithmicx}%
\usepackage{algpseudocode}%
\usepackage{listings}%
\usepackage{bm}
\usepackage{enumitem}
\usepackage{algpseudocode}
\usepackage{booktabs}
\usepackage{xurl} 
\usepackage{soul, color, xcolor}

\theoremstyle{thmstyleone}%
\theoremstyle{thmstyletwo}%

\theoremstyle{thmstylethree}%

\begin{document}

\title[Article Title]{A Regularized Block Diagonal RLS Algorithm for Acoustic Echo Cancellation}

\author[1]{\fnm{Ruibin} \sur{Hou}}\email{sqhrb@outlook.com}
\author*[1]{\fnm{Chenggang} \sur{Zhang}}\email{zhangcg@imun.edu.cn}
\author[1]{\fnm{Yufeng} \sur{Diao}}\email{diaoyufeng@imun.edu.cn}


\affil*[1]{\orgdiv{The College of Computer Science and Technology}, \orgname{ Inner Mongolia Minzu University}, \orgaddress{\city{Tongliao}, \postcode{028000}, \state{Inner Mongolia}, \country{China}}}


\abstract{While the recursive least square (RLS) algorithm is widely used in adaptive filtering applications like acoustic echo cancellation (AEC) due to its fast convergence rate, its high computational complexity severely limit its practical deployment for long filters. In this paper, a regularized block-diagonal RLS (RBD-RLS) algorithm is proposed to address these challenges. By approximating the inverse covariance matrix as a block-diagonal structure, RBD-RLS simplifies the update process into independent parallel computations of sub-blocks, effectively reducing the computational complexity. Additionally, Tikhonov regularization is applied to each sub-blocks for enhance numerical stability. A series of experimental results demonstrate that RBD-RLS maintains good convergence while significantly reducing computational complexity. Moreover, it still exhibits relative robustness in real-world scenarios.}




\maketitle

\section{Introduction}\label{sec1}

In recent years, voice interaction scenarios, such as audio/video conferencing, smart homes, and in-vehicle control systems, have become increasingly prevalent \cite{risteaICASSP2024Speech2025}
\cite{liASigmoidAffineProjection2025}
\cite{huangAffineprojectionLorentzianAlgorithm2021}
\cite{gannotConsolidatedPerspectiveMultimicrophone2017}. 
In these sensor systems, the microphone that captures voice inevitably suffers from acoustic echo, which severely degrades the quality of the sensed audio signal
\cite{gierlichCarHandsFreeTesting2014}\cite{zhangImprovingAcousticEcho2024}. 

Acoustic echo cancellation (AEC) technology is applied to subtract it from the microphone signal by estimating the echo path with adaptive filtering algorithms
\cite{jiangSmallfootprintAcousticEcho2025} 
\cite{breiningAcousticEchoControl1999} 
\cite{zhangRobustCascadedAcoustic2020}. 
However, designing effective adaptive filters requires a strict trade-off among convergence rate, steady-state error, and computational complexity.

Various algorithms in AEC have been proposed in the past decades, notably the least mean squares (LMS) algorithm families \cite{paleologuOverviewOptimizedNLMS2015a}
\cite{wangNewEffectiveNonparametric2023}
\cite{shamsiRobustDiffusionLMS2026}. They have been widely used due to the low complexity of $O(N) $ and the ease of implementation. However, when the input signal is highly correlated (eg.,speech), the convergence is relatively slow \cite{lopesAnalysisLMSNLMS2023}\cite{chenOptimizingAcousticEcho2024}. In contrast, the recursive least squares (RLS) algorithm utilizes the second-order statistical properties of the input signal to achieve an optimal convergence rate\cite{xuSimplifiedRLSAlgorithm2010} \cite{dogariuIdentificationRoomAcoustic2022}\cite{otopeleanuRobustDataReuseRegularized2025}, but it is difficult to implement in real time due to the high algorithmic complexity of $O(N^2)$ 
 and numerical instability\cite{elisei-iliescuRecursiveLeastSquaresAlgorithms2019}
\cite{yadavStateoftheartSurveyNoise2025}\cite{otopeleanuPracticalRegularizedRecursive}.

These problems stem from the recursive computation of the inverse correlation matrix, which has attracted considerable research interest in avoiding direct matrix inversion to reduce complexity\cite{zakharovLowComplexityRLSAlgorithms2008a}
\cite{cioffiFastRecursiveleastsquaresTransversal1984} \cite{sutcliffedemoraesFasterRLSDCDAdaptive2024} \cite{gouveiaNumericallyStableHouseholderBased2026}. It is worth noting that the fast RLS (FRLS) algorithm, exploiting the shifted structure of input data vectors, successfully reduces computational complexity to $O(N)$\cite{cioffiFastRecursiveleastsquaresTransversal1984}. However, FRLS suffers from severe numerical instability in finite precision implementation. Another approach is the use of iterative line search methods, notably the dichotomous coordinate descent (DCD) based RLS algorithm. The RLS-DCD algorithm solves auxiliary normal equations via binary bit-shifts, which provides excellent numerical robustness\cite{zakharovLowComplexityRLSAlgorithms2008a}. Nevertheless, its matrix update process implicitly requires $O(N^2)$ data copy operations\cite{sutcliffedemoraesFasterRLSDCDAdaptive2024} . Although this problem can be solved by using memory address modification instructions on FPGAs, the innate algorithmic complexity remains unreduced. For general-purpose processors or resource-constrained devices, the execution delay of RLS-DCD is still prohibitive.

In this article, a regularized block-diagonal RLS (RBD-RLS) algorithm is proposed to address the issue of complex matrix calculations in the RLS algorithm. By approximating the $N$-th order  autocorrelation matrix as a block-diagonal structure, it decomposes large-scale matrix updates into the parallel computation of small sub-blocks, significantly reducing the complexity. Furthermore, Tikhonov regularization\cite{douFilteringTikhonovRegularizationInversion2019}\cite{gerthNewInterpretationTikhonov2021} is applied to each sub-block, which is utilized to mitigate the ill-conditioning of the matrices during the initial adaptation stage, thereby preventing early numerical divergence.

Following this introduction, the organization of this paper is
as follows. Section \ref{sec:PD} summarizes the main task of AEC and the related work of RLS. RBD-RLS is derived in Section \ref{sec:RBD}, which involves the block diagonal approximation and matrix regularization. In Section \ref{sec:AA}, the analysis of complexity and numerical instability is presented.
The results of the experiments are provided in Section \ref{sec:exp}. Finally, Section VI concludes this work.

\section{Problem Definition and Relate Work}
\label{sec:PD}
\subsection{Acoustic Echo Cancellation System}
\begin{figure}[h]
    \centering
    \includegraphics[scale=1]{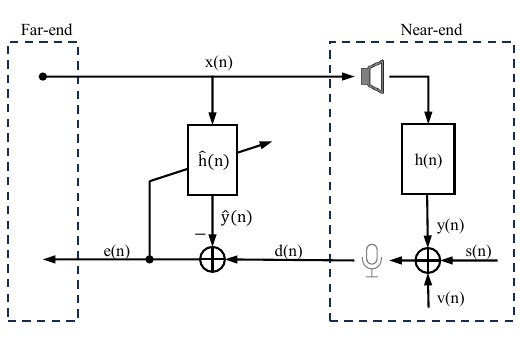}
     \caption{General framework of the AEC system.}
     \label{fig1}
\end{figure}

The general framework of the AEC system is depicted in Fig.\ref{fig1}. A voice communication system can be divided into the far-end and the near-end\cite{saremiAcousticEchoCanceller2023}. The far-end signal $x(n)$ is transmitted to the near-end loudspeaker for playback. Following acoustic reflection within the room, the echo signal $y(n)$ is formed. This process can be expressed as
\begin{equation}y(n)={h}(n)*x(n)\label{eq1}\end{equation}

Where $h(n)$ is the room impulse response (RIR), * denotes the convolution, and $n$ is the sample index. The echo $y(n)$, the background noise $v(n)$ and the near-end speech signal $s(n)$ are jointly captured by the near-end microphone to generate the mixed signal $d(n)$
\begin{equation}d(n)=s(n)+y(n)+v(n)\label{eq2}\end{equation}

The main task of AEC is to estimate ${h(n)}$ by dynamically updating adaptive filters $\hat{h}(n)$, and generating the estimated echo $\hat{y}(n)$
\begin{equation}\hat{y}(n)=\hat{{h}}(n)x(n)\label{eq3}\end{equation}

$\hat{y}(n)$ is then subtracted from $d(n)$ to obtain the error signal $e(n)$ 
\begin{equation}e(n)=d(n)-\hat{y}(n)\label{eq4}\end{equation}

When $e(n)$ is minimized, adaptive filters is considered to be convergent. Under ideal conditions, the system output is equal to the sum of the near-end speech and noise, i.e, $e(n)=s(n)+v(n)$.

\subsection{Recursive Least Squares Algorithm}
The RLS algorithm recursively updates the filter, which effectively utilizes the second-order statistical properties of the input signal to minimize the sum of weighted error squares, thereby achieving a fast convergence rate.

The signal $\textbf{x}(n)$ could seem like $[x(n),x(n-1),\dots,x(n-N+1)]^T$, where $N$ is the length of the coefficients and the filter could be $\textbf{w}(n)= [w_0,w_1,…,w_{N-1}]$.

The RLS cost function is defined as
\begin{equation}J(n)=\sum^{n}_{i=1}\lambda^{n-i}\left|d(i)-\textbf{w}^T(i)\textbf{x}(i)\right|^2\label{eq5}\end{equation}

Where $\lambda (0<\lambda\le1)$ is the forgetting factor used to control the weight of historical data. As $\lambda$ approaches $1$, RLS maintains better steady-state performance. When $\lambda$ approaches $0$, it will respond more rapidly to new data, resulting in a good tracking capability.

By taking the derivative to $\textbf{w}(n)$, the following equation is obtained
\begin{equation}\textbf{w}(n)=\textbf{R}_{xx}^{-1}(n)\textbf{R}_{xd}(n)\label{eq6}\end{equation}

Where $\textbf{R}_{xx} (n)$ and $\textbf{R}_{xd} (n)$ represent the autocorrelation matrix and the weighted cross-correlation matrix of $\textbf{x}(n)$. $\textbf{R}_{xx} (n)$ can be obtained recursively
\begin{equation}\textbf{R}_{xx}=\lambda\textbf{R}_{xx}(n-1)+\textbf{x}(n)\textbf{x}^T(n)\label{eq7}\end{equation}

The dimension of $\textbf{R}_{xx} (n)$ is determined by the filter order $N$, which could be set to thousands in practical applications. The matrix inversion lemma\cite{tylavskyGeneralizationMatrixInversion1986} is applied for calculate $\textbf{R}_{xx}^{-1}(n)$
\begin{equation}(A+BCD)^{-1}=A^{-1}-A^{-1}B(C^{-1}+DA^{-1}B)^{-1}DA^{-1}\label{eq8}\end{equation}

Defining $\textbf{P}(n)=\textbf{R}_{xx}^{-1}(n)$ and substituting it into \eqref{eq8}, the recursive update formula for \textbf{P}(n) is derived.

Consequently, the complete process of RLS consists of the following four steps
\begin{enumerate}[label={\small\itshape step\arabic*.}, leftmargin=*]
\item Calculate the gain vector $\bm{k}(n)$
    \begin{equation} \label{eq9}
        \bm{k}(n) = \frac{\bm{P}(n-1)\bm{x}(n)}{\lambda + \bm{x}^T(n)\bm{P}(n-1)\bm{x}(n)}
    \end{equation}
    
    \item Calculate the prior error $e(n)$
    \begin{equation} \label{eq10}
        e(n) = d(n) - \bm{w}^T(n-1)\bm{x}(n)
    \end{equation}
    
    \item Update the filter vector $\bm{w}(n)$
    \begin{equation} \label{eq11}
        \bm{w}(n) = \bm{w}(n-1) + \bm{k}(n)e(n)
    \end{equation}
    
    \item Update the matrix $\bm{P}(n)$
    \begin{equation} \label{eq12}
        \bm{P}(n) = \lambda^{-1}[\bm{P}(n-1) - \bm{k}(n)\bm{x}^T(n)\bm{P}(n-1)]
    \end{equation}
\end{enumerate}

Despite its fast convergence and excellent tracking performance, the standard RLS has two major drawbacks. First, its computational complexity is extremely high. $\bm{k}(n)\bm{x}^T(n)\bm{P}(n-1)$ in \eqref{eq12} requires $O(N^2)$ operations. Since filter order $N$ typically ranges from $512$ to $2048$ in practical applications, which imposes a heavy computational burden on resource-constrained devices. Second, RLS suffers from numerical instability. Because the recursive update of $\textbf{P}(n)$ involves subtraction, its positive definiteness is easily compromised by accumulated quantization and round-off errors in finite-precision arithmetic. Once $\textbf{P}(n)$ loses its positive definiteness, the term $\textbf{x}^T(n)\textbf{P}(n-1)\textbf{x}(n)$ in \eqref{eq8} approaches zero, which drives $\textbf{k}(n)$ to diverge, ultimately causing the algorithm to fail.

\section{Regularized Block Diagonal Algorithm}
\label{sec:RBD}
\subsection{Block Diagonal Approximation}
In order to address the issue of excessively high computational complexity caused by a too large value of $N$, block-processing techniques have been widely introduced into the adaptive filter algorithm\cite{clarkBlockImplementationAdaptive1981}\cite{fayadhEnhancementThreeCombining2014}.

From an estimation-theoretic perspective, both the LMS and RLS families share the ultimate objective of seeking the minimum mean-square error (MMSE) Wiener solution, which supports the extension of the block-processing paradigm to the RLS framework\cite{fayadhEnhancementThreeCombining2014}. Consequently, $\mathbf{x}(n)$ and $\mathbf{w}(n)$ are structured into sub-vectors $M$ of dimension $L \times 1$ (where $N = ML$):
\begin{equation}\textbf{x}(n)=[\textbf{x}^T_1(n),\textbf{x}^T_2(n),\dots,\textbf{x}^T_M(n)]\label{eq13}\end{equation}
\begin{equation}\textbf{w}(n)=[\textbf{w}^T_1(n),\textbf{w}^T_2(n),\dots,\textbf{w}^T_M(n)]\label{eq14}\end{equation}

The underlying mathematical justification for the covariance matrix $\mathbf{R}_{xx}(n)$ and $\textbf{P}(n)$ lies in the physical autocorrelation properties of acoustic signals, which inherently exhibit strong short-term temporal stationarity, characterized by high correlation across localized time lags that rapidly dissipates over larger time horizons. In the block-partitioned formulation, the diagonal sub-matrices of $\mathbf{R}_{xx}(n)$ intrinsically capture these dominant short-term autocorrelation features within localized time windows of length $L$. Conversely, the off-diagonal blocks represent weak cross-correlations between temporally distant signal segments, which yield diminishing returns for adaptive estimation gain.

Therefore, it can  be simplified that the matrix $\textbf{R}(n)$ is a diagonal form
\begin{equation}\textbf{R}(n)\approx diag[\textbf{R}_1(n),\textbf{R}_2(n),\dots,\textbf{R}_M(n)]\label{eq15}\end{equation}

$\textbf{R}_i (n)$ represents the $L \times L$ sub-diagonal matrix of $\textbf{R}(n)$, and $\textbf{P}(n)$ can be approximated identically.
\begin{equation}\textbf{P}(n)=\textbf{R}^{-1}(n)=diag[\textbf{P}_1(n),\dots,\textbf{P}_M(n)]\label{eq16}\end{equation}

The update of $\textbf{P}(n)$ is transformed into $M$ independent updates of $L\times L$ block matrices, significantly reducing  computational complexity.

Following the partition of $\textbf{P}(n)$ and $\textbf{x}(n)$ , the term $\textbf{P}(n-1)\textbf{x}(n)$ in \eqref{eq9} becomes a vector composed of $M$ sub-blocks. Define the temporary variable $\textbf{v}_i (n)=\textbf{P}_i(n-1)\textbf{x}_i (n)$
\begin{equation}
\begin{split}
\bm{P}(n-1)\bm{x}(n) &\approx 
\begin{bmatrix}
\bm{P}_1 &        & \\
         & \ddots & \\
         &        & \bm{P}_M
\end{bmatrix}
\begin{bmatrix}
\bm{x}_1 \\
\vdots \\
\bm{x}_M
\end{bmatrix}\\
&=
\begin{bmatrix}
\bm{P}_1 \bm{x}_1 \\
\vdots \\
\bm{P}_M \bm{x}_M
\end{bmatrix}
=
\begin{bmatrix}
\bm{v}_1(n) \\
\vdots \\
\bm{v}_M(n)
\end{bmatrix}
\end{split}
\end{equation} \label{eq17}

Define the scalar $g(n)$ to represent the inner product
\begin{equation}
\begin{split}
g(n) &=\textbf{x}^T(n)\textbf{P}(n-1)\textbf{x}(n)\\
&=[\textbf{x}^T_1(n) \cdots \textbf{x}^T_M(n)]
\begin{bmatrix}
\bm{v}_1 \\
\vdots    \\
\bm{v}_M(n)
\end{bmatrix}
=
\sum_{i=1}^M\textbf{x}^T_i(n)\textbf{v}_i(n)
\end{split}
\end{equation} \label{eq18}

Substituting $g(n)$ into \eqref{eq9}, the denominator is defined as a global normalization factor $D_{inv}(n)$ 
\begin{equation}
D_{inv}(n)=\frac{1}{\lambda +g(n)}
\end{equation} \label{eq19}

The calculation of $\bm k(n)$ is thus simplified to scalar multiplication over a block matrix, and it can be divided as follows
\begin{equation}
\textbf{k}(n)=
\begin{bmatrix}
    \textbf{v}_1(n) \\
    \vdots \\
    \textbf{v}_M(n)
\end{bmatrix}
\odot D_{inv}(n)=
\begin{bmatrix}
    \textbf{k}_1(n) \\
    \vdots\\
    \textbf{k}_M(n)
\end{bmatrix}
\end{equation} \label{eq20}

Whereas the matrices $\textbf{P}_i (n)$ are decoupled under the block-diagonal assumption,  $\textbf{k}_i (n)$ remains coupled through $D_{inv} (n)$ and $\textbf{x}(n)$.

$\textbf{w}(n)$ and $e(n)$ can be processed in blocks identically. 
\begin{equation}
y_{priori}(n)=\textbf{w}^T(n-1)\textbf{x}(n)=\sum_{i=1}^M\textbf{w}^T_i(n-1)\textbf{x}_i(n)
\end{equation} \label{eq21}

\begin{equation}
e(n)=d(n)-y_{priori}(n)
\end{equation} \label{eq22}

The update of $\textbf{w}(n)$ is executed independently for each sub-block
\begin{equation}
\textbf{w}_i(n)=\textbf{w}_i(n-1)+\textbf{k}_i(n)e(n)
\end{equation} \label{eq23}

Similarly, the update of $\bm P(n)$ is performed independently across sub-blocks
\begin{equation}
\begin{split}
\textbf{P}_i(n)&=\frac{1}{\lambda}[\textbf{P}_i(n-1)-\textbf{k}_i(n)\textbf{x}_i^T(n)\textbf{P}_i(n-1)]\\
&=\frac{1}{\lambda}[\textbf{P}_i(n-1)-\textbf{k}_i(n)\textbf{v}_i^T(n)]
\end{split}
\end{equation} \label{eq24}

\subsection{Matrix Regularization}
Although the preceding derivation decomposes complex matrix calculations into $M$ lower-order diagonal matrices, numerical instability remains a risk during the $\textbf{P}(n)$ update. Therefore, Tikhonov regularization (diagonal loading) is incorporated to guarantee numerical stability. The cost function is modified to the following
\begin{equation}
    J_{req}(n)=\sum_{i=1}^n\lambda^{n-i}[d(i)-\textbf{w}^T(n)\textbf{x}(i)]^2+\delta||\textbf{w}(n)||^2
\end{equation}\label{eq25}

The regularization parameter $\delta \textbf{I}$ is a small positive constant. The modification is equivalent to adding $\delta \textbf{I}$ to $\textbf{R}_{xx}(n)$, which produces $\textbf{R}_{reg}(n)=\textbf{R}_{xx}(n)+\delta \textbf{I}$. Consequently, the initialization of \textbf{R}(0) shifts from 0 to $\delta \textbf{I}$. And the initialization of the inverse matrix \textbf{P}(n) becomes
\begin{equation}
    \textbf{P}(0)=\textbf{R}_{reg}^{-1}(0)=\delta ^{-1}\textbf{I}
\end{equation}\label{eq26}

By regularizing each sub-block of \textbf{P}(n), the initialization steps for the proposed RBD-RLS are formulated as
\begin{equation}
    \textbf{w}_i(0)=0 (i=1 \cdots M)
\end{equation}\label{eq27}
\begin{equation}
    \textbf{P}_i(0)=\delta^{-1}\textbf{I}_L (i=1 \cdots M)
\end{equation}\label{eq28}
Where $\textbf{I}_L$ denotes the identity matrix of L×L.
\section{Algorithm Analysis}
\label{sec:AA}
This section provides a further theoretical analysis of RBD-RLS. Based on mathematical derivations, the pseudocode of the RBD-RLS is shown as :

\begin{algorithm}[htbp!]
\caption{The regularized block-diagonal RLS}\label{algo}
\begin{algorithmic}[1]
\Require $N, L, M=N/L, \lambda, \delta$

\State \textbf{Initialization:}
\State $P_i \leftarrow \delta^{-1}I_L$
\State $y_{priori} \leftarrow 0$
\State $g_{sum} \leftarrow 0$

\State \textbf{Pre-computation:}
\For{$i = 1 \to M$}
    \State $x_i \leftarrow x_{buf}((i-1)L+1:iL)$
    \State $v_i \leftarrow P_i x_i$
    \State $g_i \leftarrow x_i^T v_i$
    \State $g_{sum} \leftarrow g_{sum} + g_i$
    \State $y_{priori} \leftarrow y_{priori} + W_i^T x_i$
    \State $v_{blocks}(i,:) \leftarrow v_i^T$
\EndFor

\State $e(n) \leftarrow d(n) - y_{priori}$
\State $D_{inv} \leftarrow 1.0 / (\lambda + g_{sum})$

\State \textbf{Parallel update sub-block:}
\For{$i = 1 \to M$}
    \State $v_i \leftarrow v_{blocks}(i,:)^T$
    \State $k_i \leftarrow v_i \cdot D_{inv}$
    \State $w_i \leftarrow w_i + k_i e(n)$
    \State $P_i \leftarrow (P_i - k_i v_i^T) / \lambda$
\EndFor
\end{algorithmic}
\end{algorithm}

\subsection{Computational Complexity Analysis}
The pseudocode reveals that RBD-RLS can be divided into three stages. For $M$ blocks, the complexity of the pre-computation and priori error calculation stage is $M\times(O(L^2)+2\cdot O(l))\approx O(ML^2)$; The complexity of the parallel sub-block update stage is $M\times(2\cdot O(L)+O(L^2 ))\approx O(ML^2)$; The complexity of the posterior output stage is $O(ML)$. Therefore, the overall complexity is the following
\begin{equation}O(ML^2 )+O(ML^2 )+O(ML)\approx O(ML^2 )=O(NL)\label{eq29}\end{equation}

Compared to the $O(N^2)$ complexity of  RLS, RBD-RLS  significantly reduces computational demand, and the number of memory copies in RBD-RLS has also decreased notably compared to RLS-DCD. The computational load of the adaptive algorithms is shown in Table \ref{tab:complexity}.

\begin{table*}[htbp]
\centering
\caption{COMPREHENSIVE COMPLEXITY OF ADAPTIVE ALGORITHM}
\label{tab:complexity}
\setlength{\tabcolsep}{3pt} 
\begin{tabular}{|p{0.3\linewidth}|c|}
\hline
Algorithm & Complexity \\ 
\hline 
NLMS                 & $O(N)$      \\
RLS                  & $O(N^2)$ \\
RLS-DCD ($N_u = 8$) & $O(N^2)$ \\
FRLS                 & $O(N)$ \\
RBD-RLS ($L = 32$)   & $O(32N)$  \\
RBD-RLS ($L = 64$)   & $O(64N)$ \\
RBD-RLS ($L = 128$)  & $O(128N)$ \\
\hline
\end{tabular}
\end{table*}

Regularization ensures that $\textbf{P}_i (0)$ is strictly positive. With the properities of  the matrix lemma \eqref{eq8}, provided that $\lambda>0$ and the normalization factor $(\lambda+g_{sum})>0$. It should be noted that Tikhonov regularization can only maintain positive definiteness in the initial stage. When the number of iterations is too large, the positive definiteness of the algorithm will deviate.

\subsection{Parameter Analysis}
Despite the reduction in computational complexity, neglect of correlation makes the rate of convergence inferior to that of RLS.

Additionally, the convergence rate of RBD-RLS is associated with the parameter $L$, the forgetting factor $\lambda$ and the regularization parameter $\delta$. As shown in Table \ref{tab:complexity}, a larger $L$ leads the convergence rate to approach that of RLS, but complexity increases. In the extreme case where $L=N$, its performance is equivalent to that of RLS, but the complexity reaches its maximum $O(N^2)$. The different input values of $\lambda$ and $\delta$ will also lead to different performances of the algorithm, which is shown in Section \ref{expa}.

\section{Experiment And Results}
\label{sec:exp}
In this paper, five experiments scenarios are designed to evaluate the performance of the proposed RBD-RLS algorithm. Section \ref{expa} investigates the impact of different forgetting factors ($\lambda$) and regularization parameters ($\delta$) on the algorithm's performance using random Gaussian white noise. 

For the subsequent four scenarios, FRLS and RLS-DCD \footnote{The code of the RLS-DCD algorithm is available at \url{https://github.com/ndemoraes/Fast-RLS-DCD-MATLAB}} are introduced for broader comparisons. Specifically, Section \ref{expb} and Section \ref{expc} employ random Gaussian white noise and colored noise as the far-end input signals, respectively. Section \ref{expd} assesses the re-convergence tracking capability when subjected to abrupt changes in the echo path. To ensure statistical reliability, the first four experiments (Section \ref{expa} to \ref{expd}) were independently executed 100 times, and their averaged results are presented. 

Finally, to verify the robustness of the proposed algorithm in real-world environments, Section \ref{expe} utilizes practical input and mixed signals sourced from the AEC challenge dataset \cite{sridharICASSP2021Acoustic2021}. 

Normalized system misalignment (MIS) is adopted as the primary evaluation metric. MIS measures the ratio of the Euclidean norm of the error vector (between $\textbf{w}$ and $\textbf{h}$) to the norm of $\textbf{h}$, expressed in decibels (dB)
\begin{equation}
    MIS = 10lg\left(\frac{\vert\vert \textbf{h}-\textbf{w}\vert\vert ^2}{\vert \vert \textbf{h}\vert \vert ^2}\right)(dB)
\end{equation}

For Section \ref{expe}, echo return loss enhancement (ERLE) is used, which defines the energy ratio between $d(n)$ and $e(n)$
\begin{equation}
    ERLE = 10lg\left(\frac{E[d^2(n)]}{E[e^2(n)]}\right)(dB)
\end{equation}

A higher ERLE value indicates better echo cancellation performance. When $ ERLE \textless 0 $, effective filtering is not achieved.

\subsection{Parameter Sensitivity Analysis}
\label{expa}
\begin{figure}[h]
    \centering
    \includegraphics[scale=1]{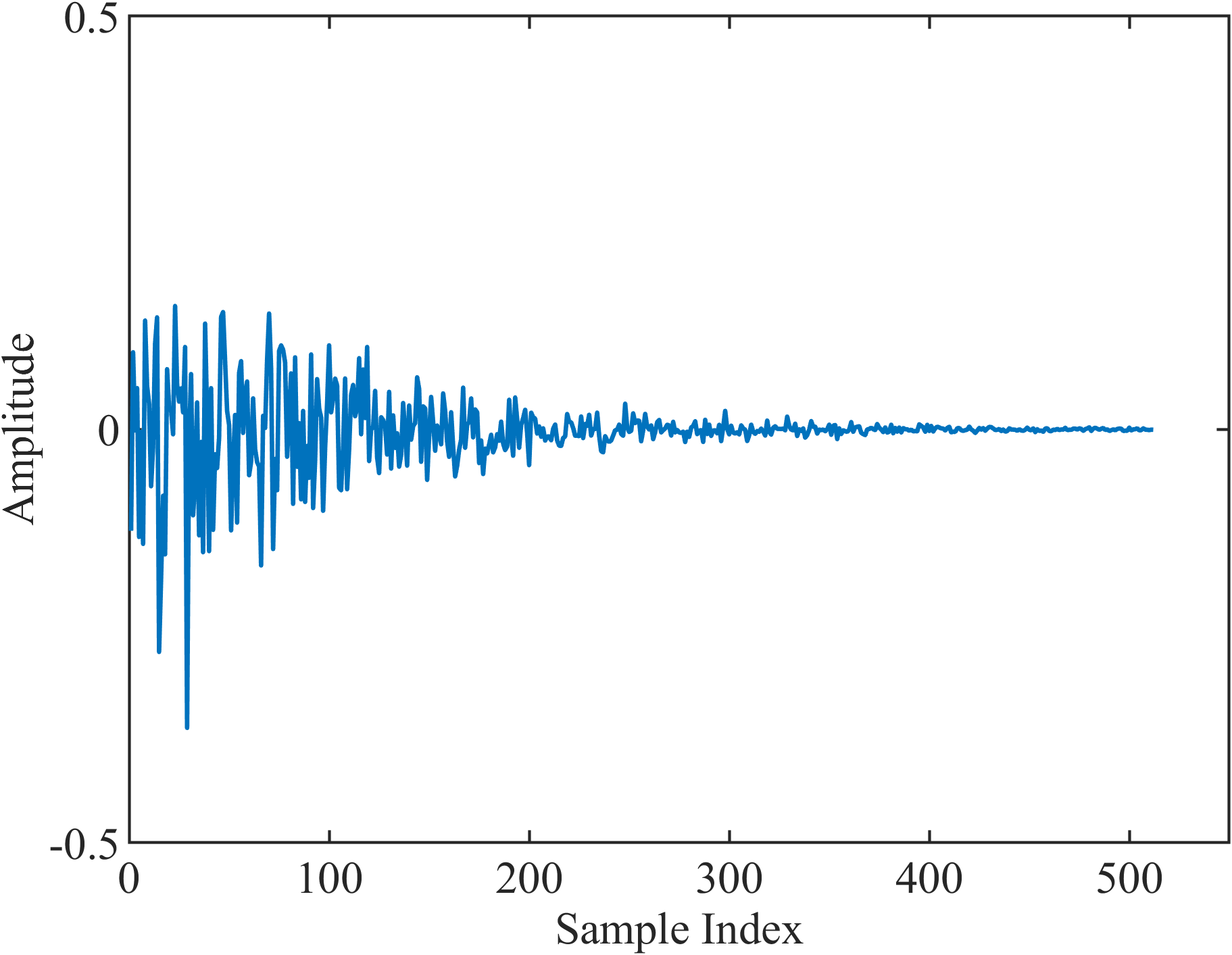}
     \caption{Random echo path.}
     \label{fig2}
\end{figure}

\begin{figure}[h]
    \centering
    \includegraphics[scale=1]{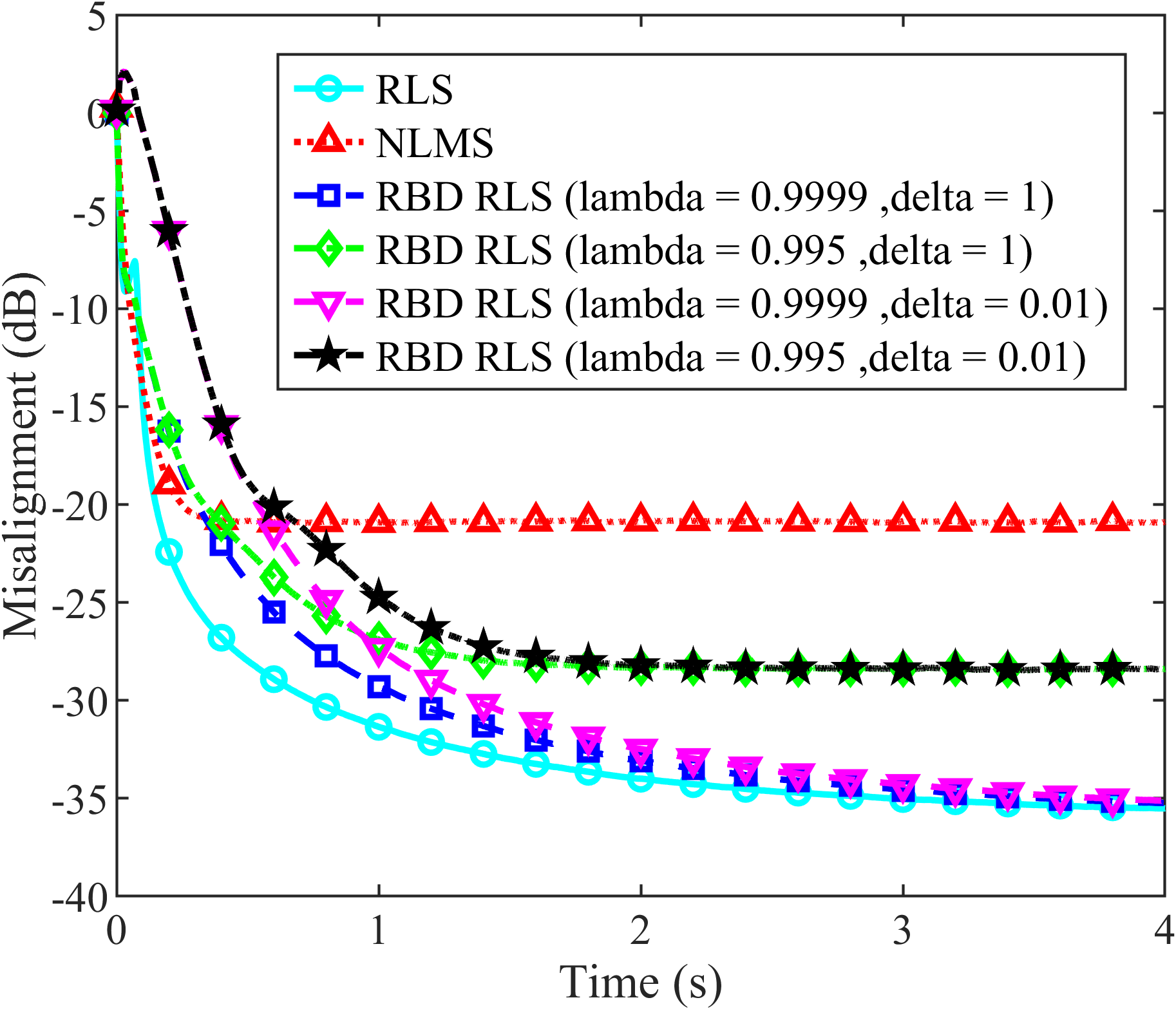}
     \caption{MIS with different value of $\delta$ and $\lambda $.}
     \label{fig3}
\end{figure}

In this experiment, four different values of $\lambda$ and $\delta$ in RBD-RLS were incorporated into this experiment. RLS and NLMS were introduced as
the comparison algorithms. The input signal $x(n)$ is generated by random white Gaussian noise of zero-mean with a signal duration of 4s and a sampling frequency of $8 kHz$. The parameters are set to $N=512$, $L=64$. The echo path is generated using a random signal in an order consistent with $N$. An attenuation factor of $0.01$ is set to simulate the signal decay process. The generated echo path is shown in Fig.\ref{fig2}. Background noise with a Signal-to-Noise Ratio (SNR) of $20$ dB is also added to $d(n)$. The results are shown in Fig.\ref{fig3}.

As illustrated in Fig.\ref{fig3}, the performance of the proposed RBD-RLS algorithm is affected by the forgetting factor $\lambda$ and the regularization parameter $\delta$. While a smaller $\lambda=0.995$ accelerates initial adaptation, it severely degrades the steady-state misalignment compared to $\lambda=0.9999$. Concurrently, an insufficient regularization parameter (e.g., $\delta=0.01$) induces an excessively aggressive initial weight update matrix ($P_{i} = \delta^{-1}I$), resulting in the severe transient overshoots observed in the early stage. By adopting the optimal configuration ($\lambda=0.9999$, $\delta=1$), the proposed algorithm effectively suppresses these early fluctuations, achieving a smooth and monotonic convergence. Notably, this optimal configuration perfectly matches the ultimate steady-state precision of the computationally demanding RLS, significantly outperforming the NLMS baseline. The marginal delay in the initial convergence rate is an acceptable trade-off for the substantial reduction in computational complexity, validating the mathematical efficacy of the block-diagonal approximation.

\subsection{Gaussian White Noise Scenario}
\label{expb}
\begin{figure}[h]
    \centering
    \includegraphics[scale=1]{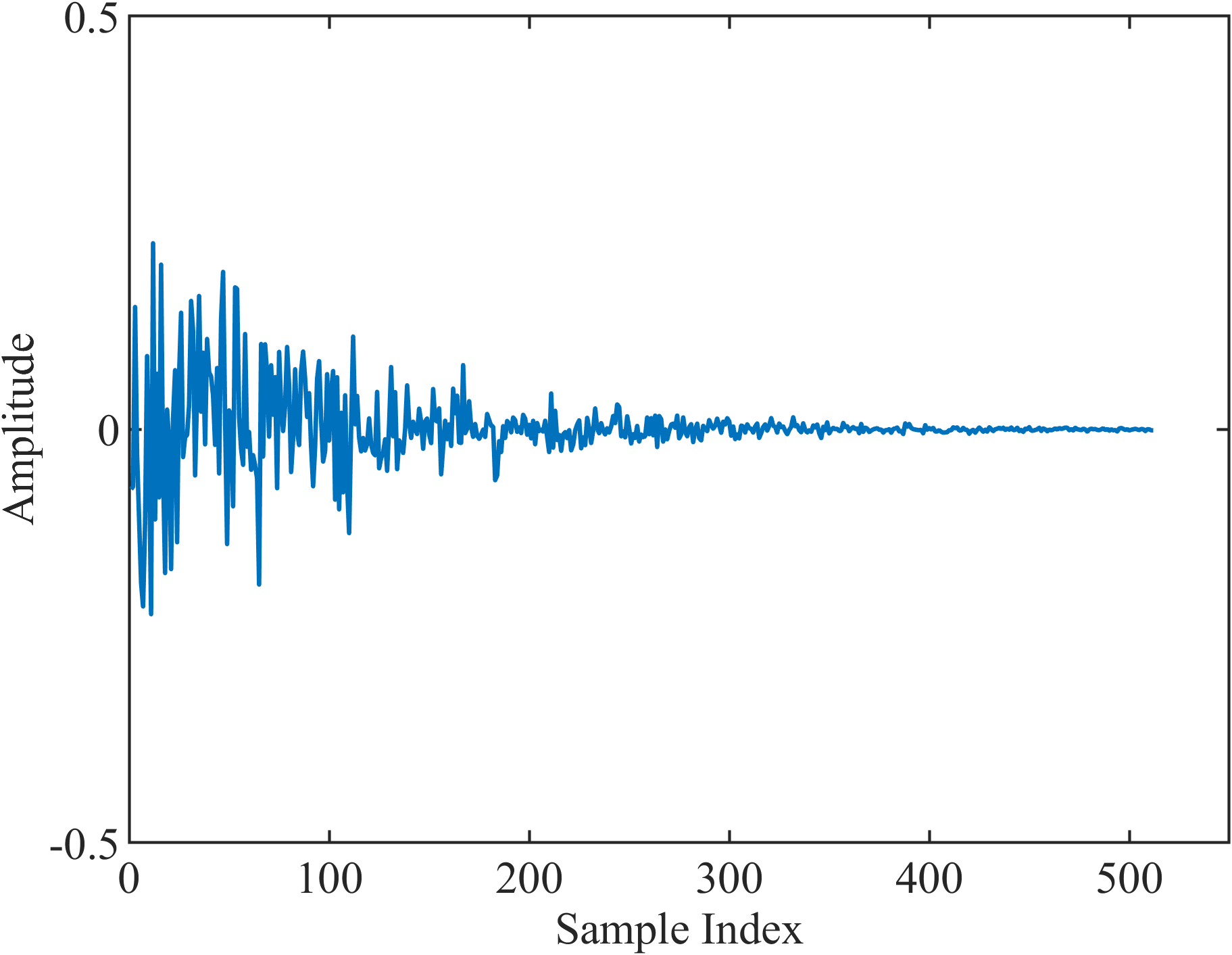}
     \caption{Random echo path.}
     \label{fig4}
\end{figure}

\begin{figure}[h]
    \centering
    \includegraphics[scale=1]{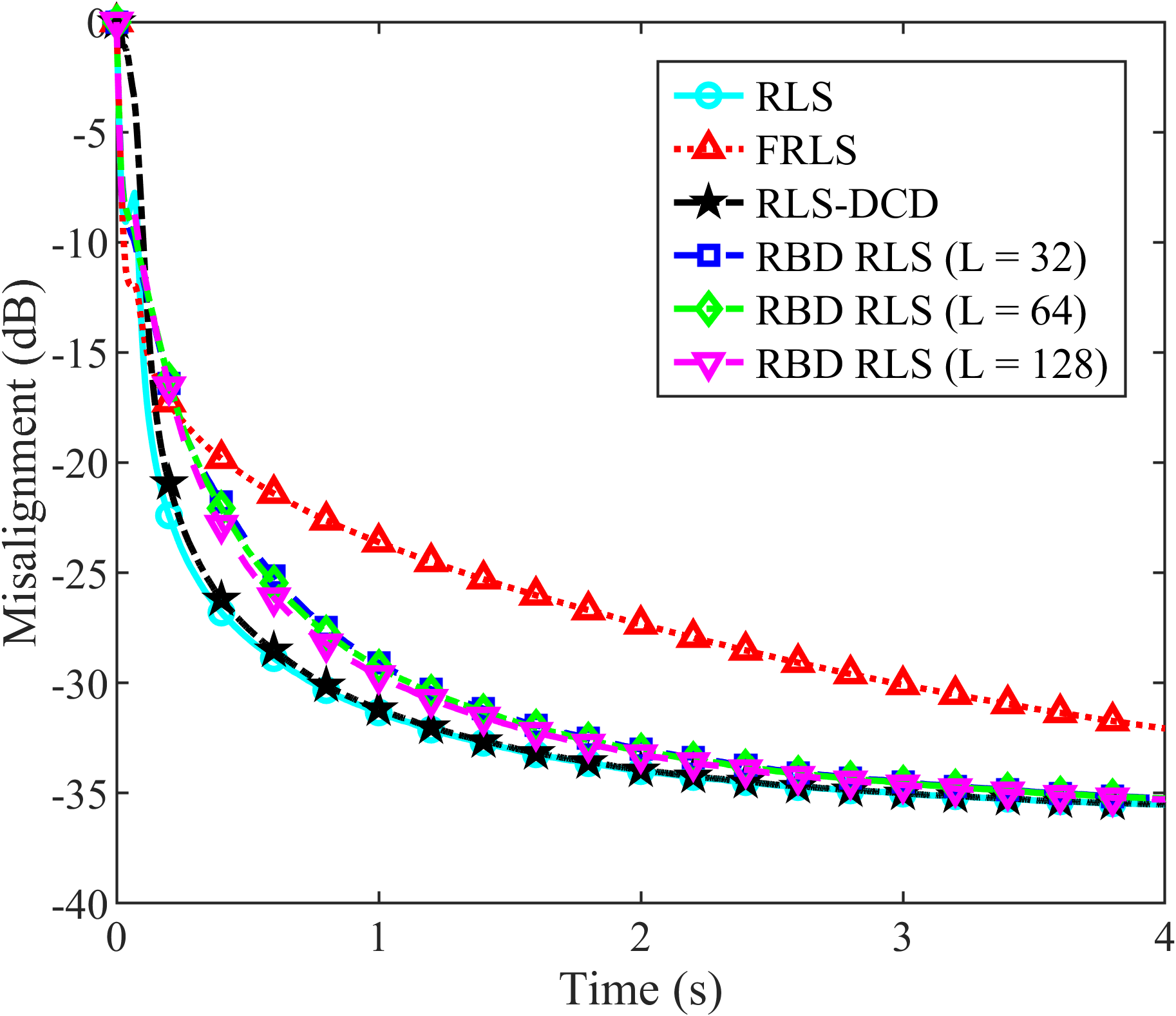}
     \caption{MIS with white noise input.}
     \label{fig5}
\end{figure}

This paper focuses solely on improvements to the RLS algorithm, so we introduced RLS, FRLS and RLS-DCD as the comparison algorithms.The input signal $x(n)$ is generated by random white Gaussian noise like Section \ref{expa}. The parameters are set to $\lambda=0.9999$, and $\delta=1$. The other parameters are the same as those in Section \ref{expa}. The echo path is shown in Fig.\ref{fig4}. The results are shown in Fig.\ref{fig5}.

It is quite obvious from Fig.\ref{fig5} that the RLS and RLS-DCD exhibit the fastest convergence rate attribute to execute the calculations of the matrix $N \times N$ in each iteration and the dichotomous coordinate
descent method. RLS and RLS-DCD achieve convergence after approximately $5000$ iterations (0.6 seconds) in the MIS of $-30$ dB, maintaining the lowest steady-state error. Although the FRLS reduces complexity to $O(N)$, it suffers from severe degradation during the initialization phase, resulting in a slow convergence rate.

By neglecting the correlation between different time instances, RBD-RLS achieves a convergence rate slightly trailing RLS while outperforming FRLS. Fig.\ref{fig5} reveals that RBD-RLS converges in 1s. In the period of 1s to 3s, the steady-state stability is demonstrably superior to the FRLS.
Due to the low autocorrelation of white noise, varying $L$ exerts a negligible influence on RBD-RLS.
\subsection{Colored Noise Scenario}
\label{expc}
\begin{figure}[h]
    \centering
    \includegraphics[scale=1]{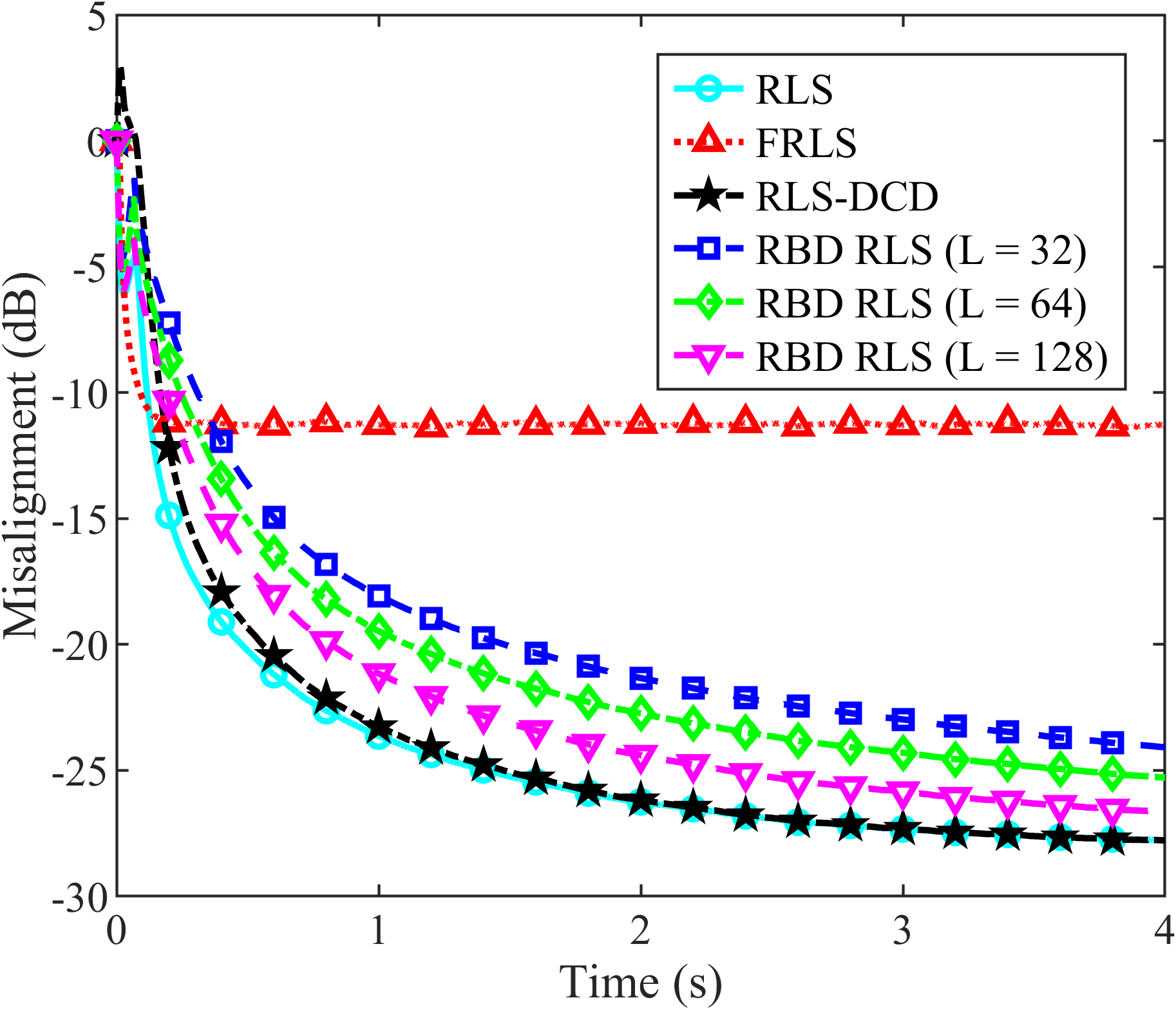}
     \caption{MIS with colored noise input.}
     \label{fig6}
\end{figure}

In this simulation, $x(n)$ was generated by passing a zero-mean Gaussian sequence through a first-order auto-regressive (AR) model with the transfer function $H(z)=\frac{1}{1-0.8z^{-1}}$, which resulted in highly autocorrelated signals. All parameters remained identical to Section \ref{expa}. The MIS are presented in Fig.\ref{fig6}.

The results demonstrate that a larger $L$ yields a stronger capability for RBD-RLS to process highly autocorrelated signals, and its convergence rate approaches that of RLS more closely.

\subsection{Echo Path Changes Scenario}
\label{expd}
\begin{figure}[h]
    \centering
    \includegraphics[scale=1]{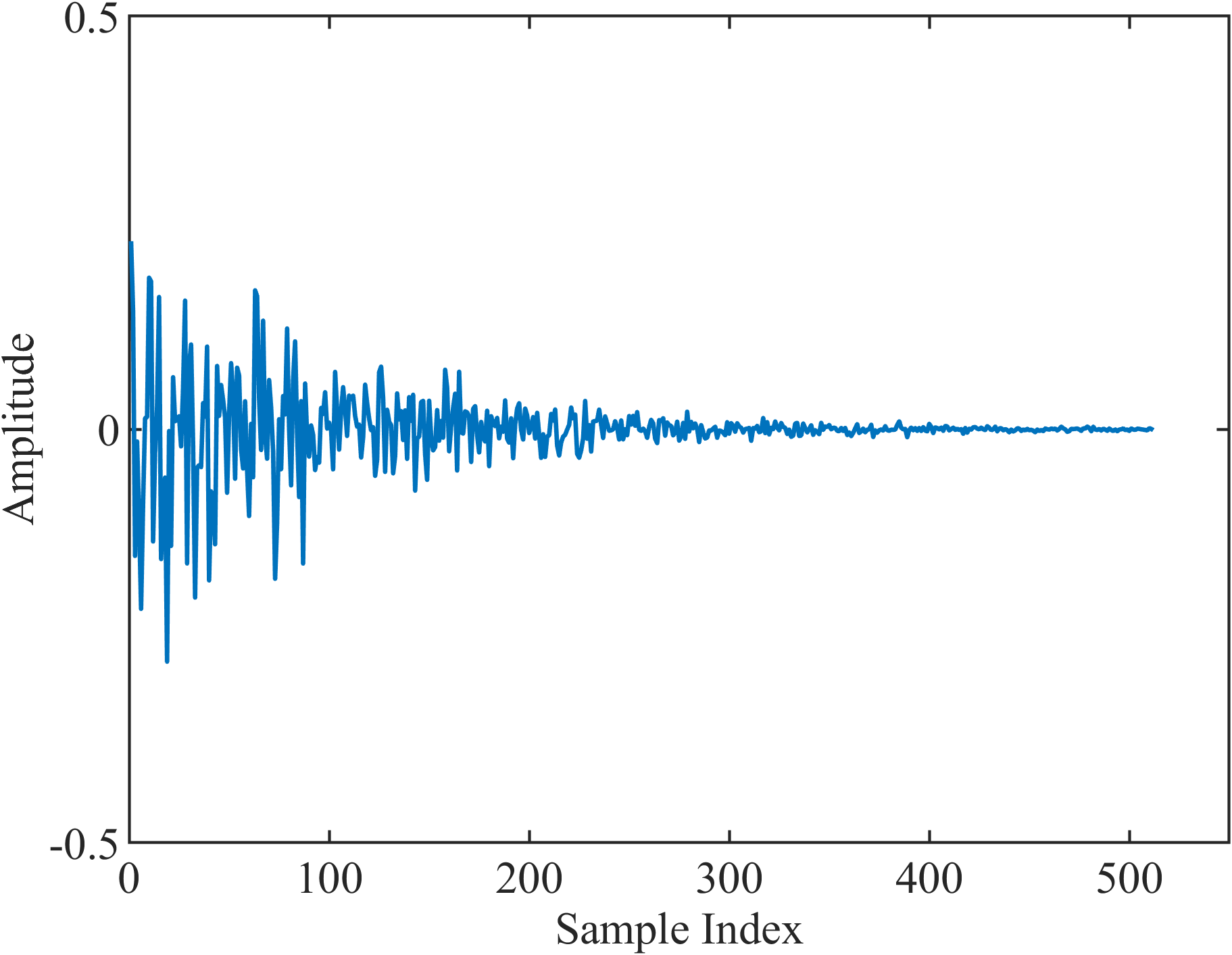}
     \caption{Echo path A.}
     \label{fig7}
\end{figure}

\begin{figure}[h]
    \centering
    \includegraphics[scale=1]{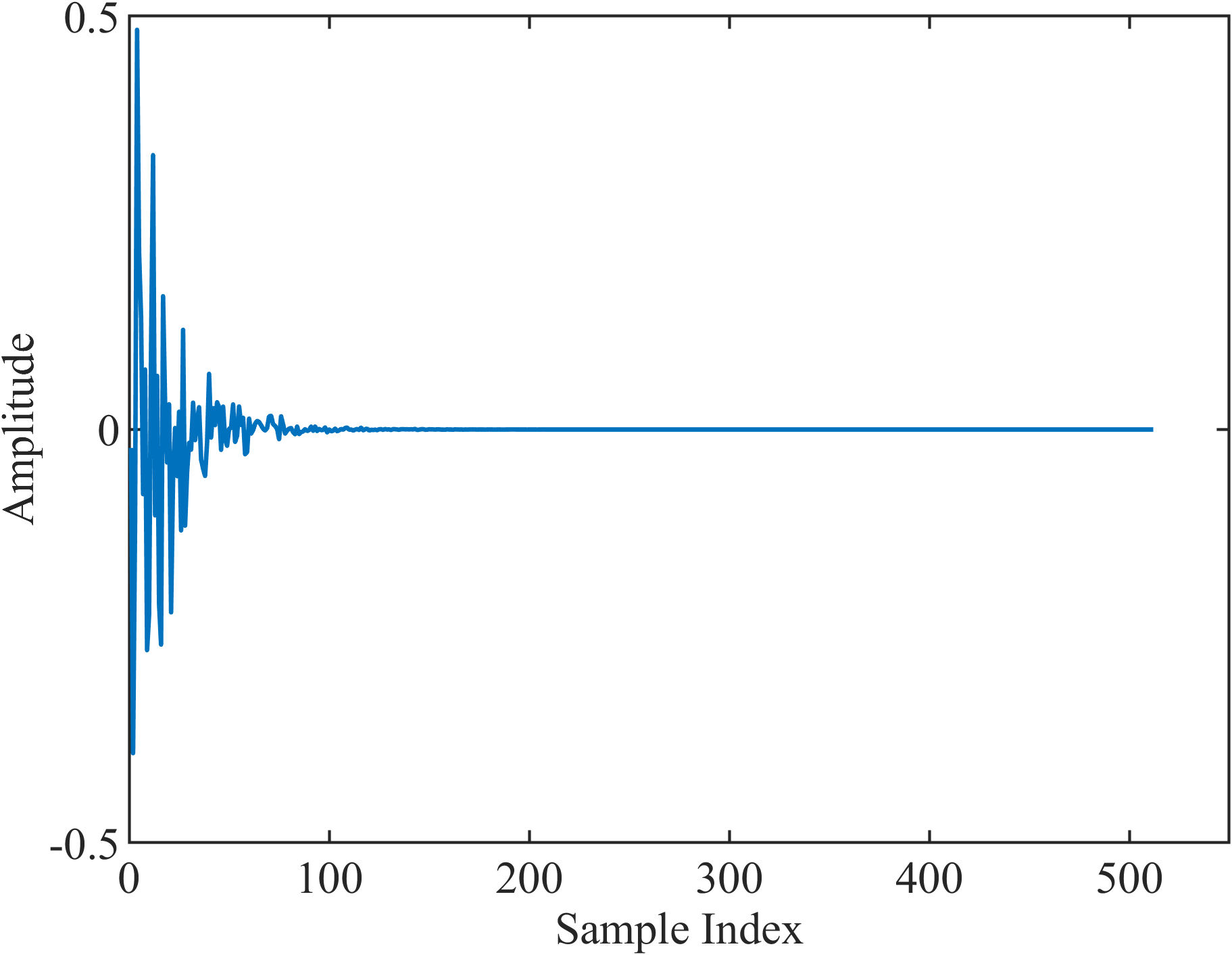}
     \caption{Echo path B.}
     \label{fig8}
\end{figure}

To verify the re-convergence tracking capabilities, two distinct echo paths (both of order $512$) were utilized. The simulation switched from echo path A to echo path B In the third second. Echo path A featured an attenuation factor of $0.01$, while path B used $0.05$. Their waveforms are shown in Fig.\ref{fig7} and Fig.\ref{fig8}, respectively. 

\begin{figure}[!th]
    \centering
    \includegraphics[scale=1]{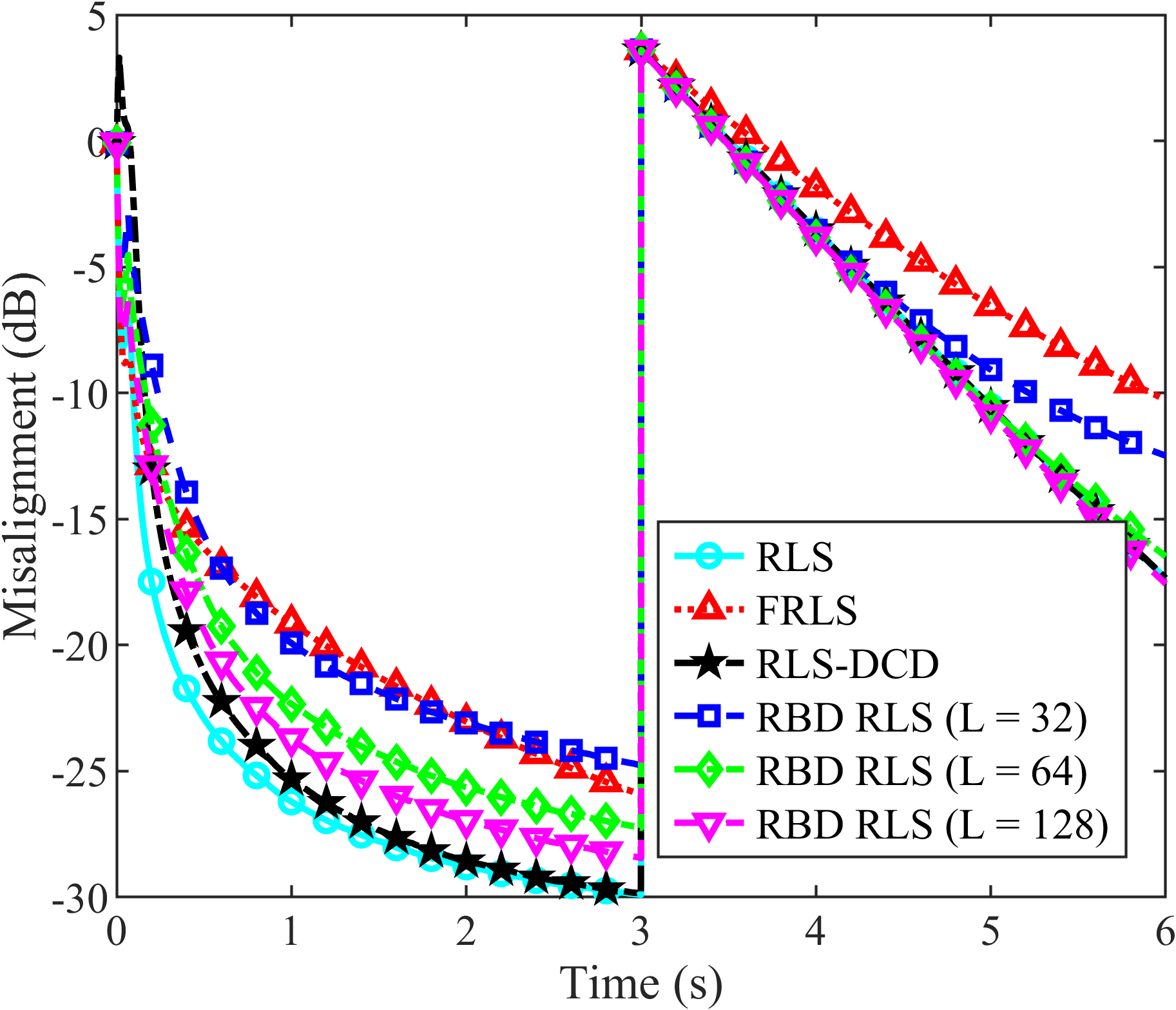}
     \caption{MIS with RIR change.}
     \label{fig9}
\end{figure}

The far-end signal was generated identically to Section \ref{expc}. Fig.\ref{fig9} illustrates the MIS tracking curves of the RBD-RLS, RLS, RLS-DCD and FRLS.

As observed, when the RIR swifts, all algorithms experience deterioration but manage to re-converge rapidly. The RLS recovers the fastest, whereas FRLS lags. In particular, when $L=128$, the proposed RBD-RLS exhibits a re-convergence trajectory close to that of RLS and RLS-DCD. Although increasing $L$ incurs a higher complexity for RBD-RLS, it strictly remains far below the $O(N^2)$ complexity of RLS, which confirms that RBD-RLS provides near-optimal re-convergence rate at a fraction of the computational expense.

\subsection{Real-World Scenario}
\label{expe}
\begin{figure}[!th]
    \centering
    \includegraphics[scale=1]{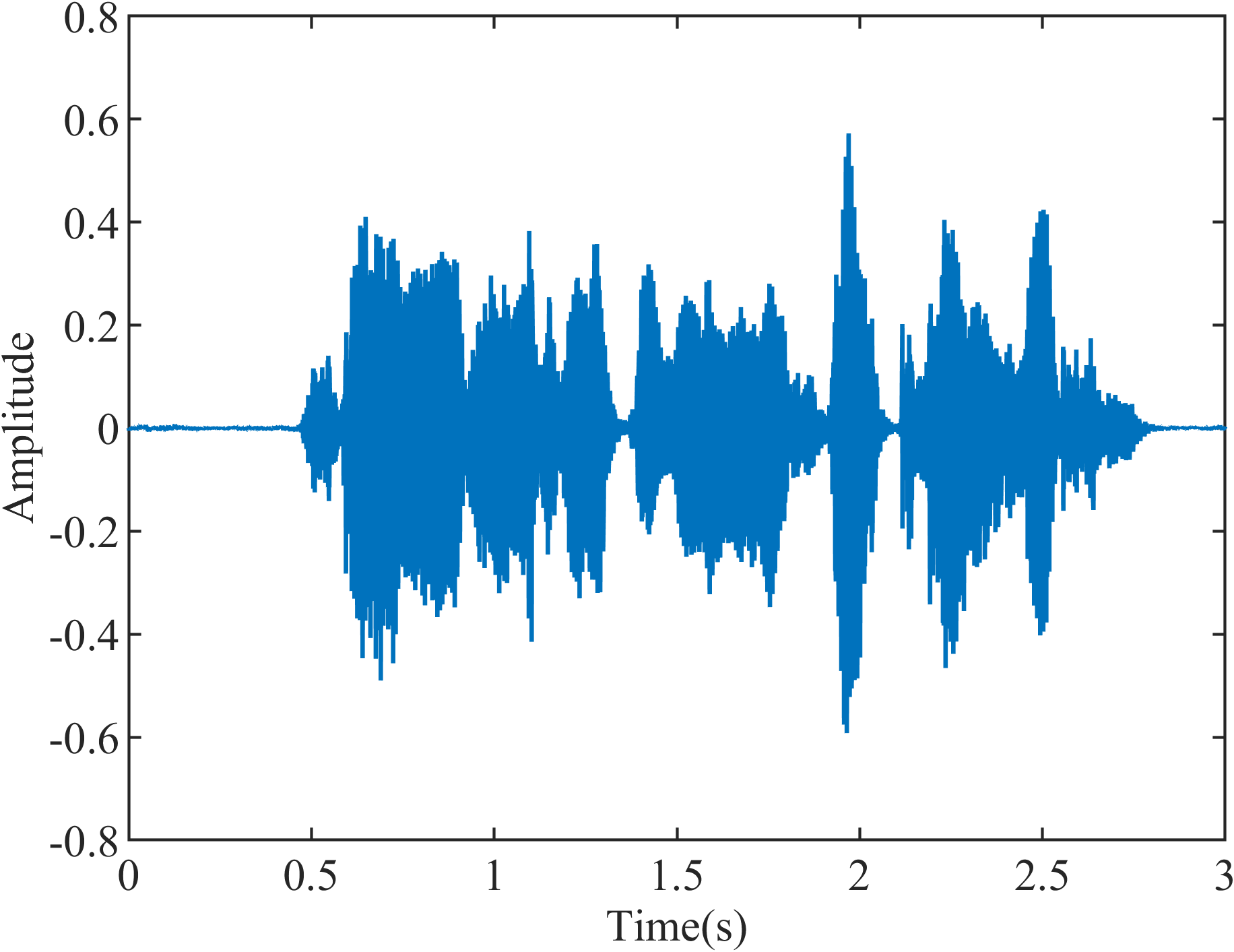}
     \caption{Far-end signal\protect\footnotemark.}
     \label{fig10}
\end{figure}

\begin{figure}[!th]
    \centering
    \includegraphics[scale=1]{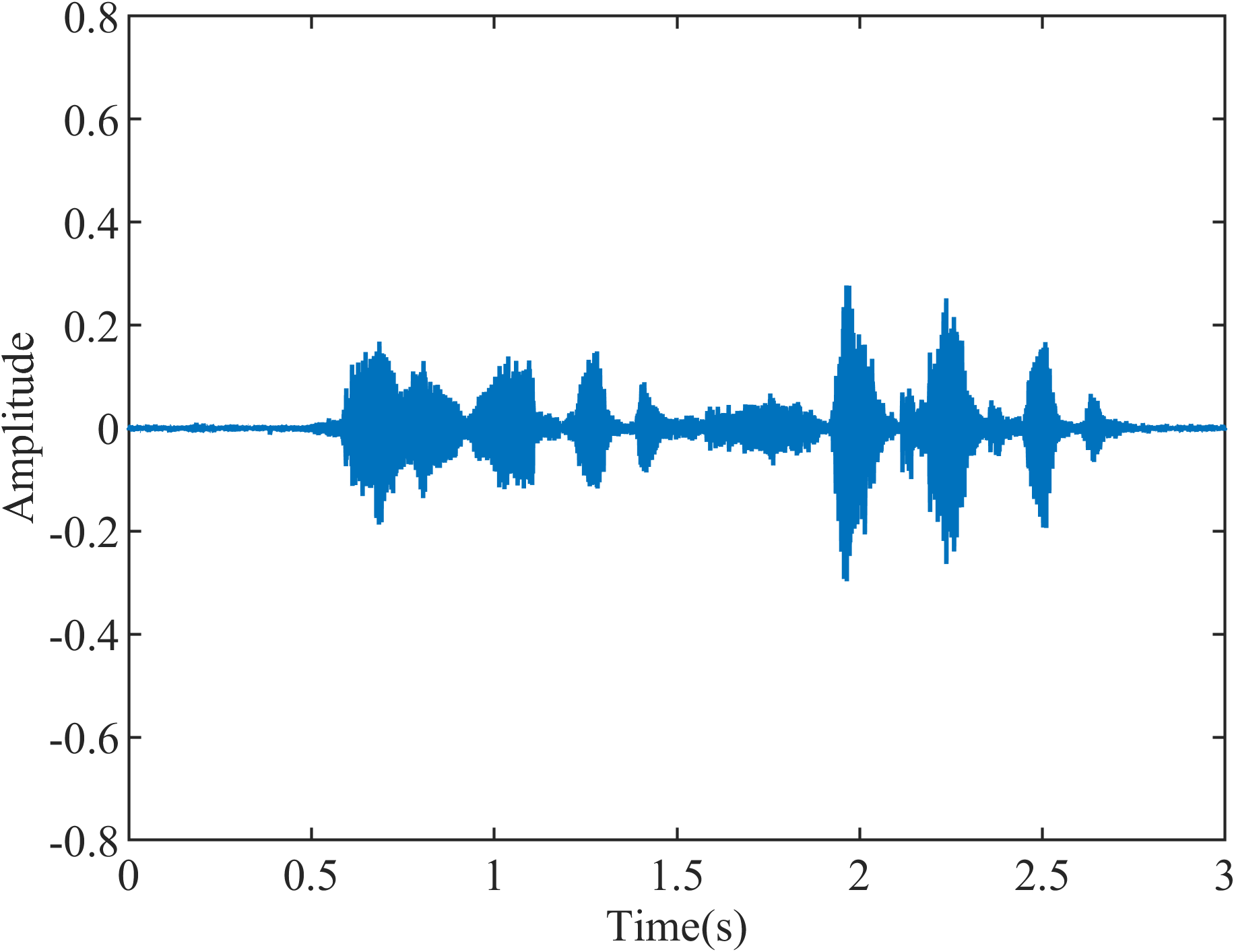}
      \caption{Microphone signal.}  
     \label{fig11}
\end{figure}

To evaluate the proposed algorithm under realistic conditions, we utilize the "noisy" blind test set from the ICASSP AEC challenge\footnote{The blind test set from the ICASSP AEC challenge is available at \url{https://github.com/microsoft/AEC-Challenge}}. This dataset which contains background noise, reverberation, and nonlinear distortions. Because these real-world recordings contain inherent time delays between the far-end and near-end paths, the generalized cross-correlation phase transform (GCC-PHAT) method is employed to estimate and align the signals prior to filtering. For the sake of visual clarity and to better highlight the transient filtering performance, the signal presented in the subsequent figures are truncated to display a specific 3-second segment from 0.5s to 3.5s. The far-end and the microphone signals are shown in Fig.\ref{fig10} and Fig.\ref{fig11}, respectively. The parameters matched those of Section \ref{expa}.

 \footnotetext{The speech signal ID is   “\path{_5z9G2AP806bhcI0QF18Qg_farend_singletalk}" }
 
\begin{figure}[!th]
    \centering
    \includegraphics[scale=1]{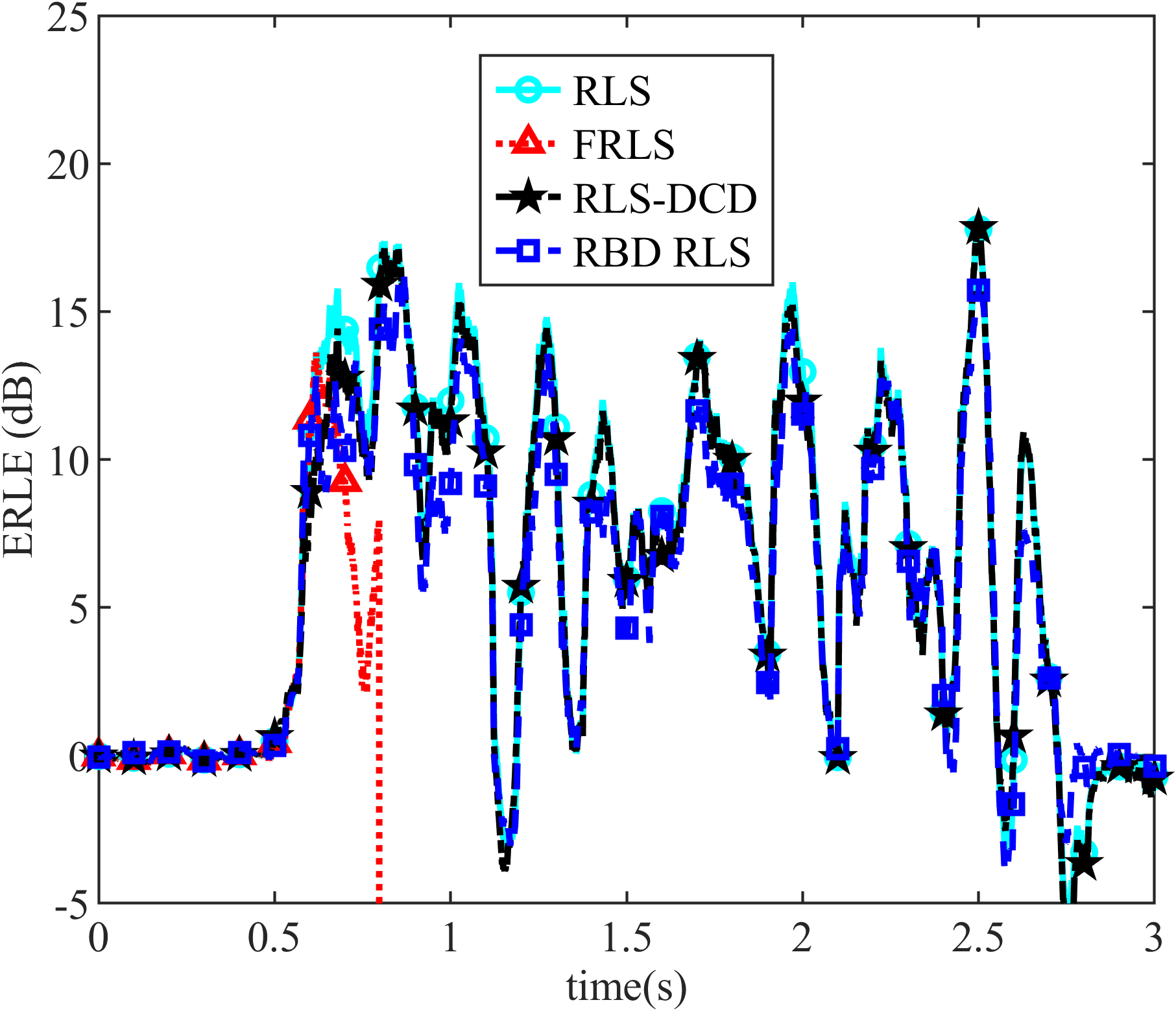}
     \caption{ERLE with real-world scenrio.}
     \label{fig12}
\end{figure}

Notably, the FRLS diverged entirely in 0.5s, failing to provide any effective filtering. Fig.\ref{fig12} plots the result of ERLE.

The result of ERLE clearly indicate that the filtering capability of the RBD-RLS practically close to that of RLS and RLS-DCD. In contrast, numerical instabilities triggered a total divergence in the FRLS around the 0.5s, rendering it unsuitable for practical AEC, which is confirmed in \cite{dogariuPerformanceDataReuseFast2022}.

The result of Fig.\ref{fig12} underscores that RBD-RLS can achieve an effect similar to that of RLS and RLS-DCD with less computational complexity, and numerical stability can be guaranteed. Due to the feature of block diagonal and matrix regularization, the proposed RBD-RLS is superior to FRLS, indicating its good robustness in real-world Scenario. 

\section{Conclusion}
\label{sec:con}
In order to solve the problem of high computational complexity of the RLS in the AEC application, a regularized block-diagonal RLS (RBD-RLS) is proposed in this paper. By approximating the $N$-th order autocorrelation matrix into $M$ sub-matrices of order $L$, the complexity is drastically reduced from $O(N^2)$ to $O(NL)$. Meanwhile, the integration of Tikhonov regularization maintains numerical stability. Extensive results confirm that, despite the significant reduction in computational load, the convergence speed of the proposed RBD-RLS remains competitive with that of the standard RLS. More importantly, it demonstrates superior robustness when processing real-world speech. Consequently, for resource-constrained AEC systems, the proposed RBD-RLS is an extremely attractive solution. Future work will expand to address advanced challenges, including double-talk detection and nonlinear echo cancellation.

\bmhead{Author contribution}
All authors contributed to the study conception and design. Material preparation, data collection and analysis were performed by Chenggang Zhang, Ruibin Hou and Yufeng Diao. The first draft of the manuscript was written by Ruibin Hou and all authors commented on previous versions of the manuscript. All authors read and approved the final manuscript.

\bmhead{Conflict of interest}

The authors declare no conflict of interests.

\bibliography{sn-bibliography}


\begin{thebibliography}{32}
\ifx \bisbn   \undefined \def \bisbn  #1{ISBN #1}\fi
\ifx \binits  \undefined \def \binits#1{#1}\fi
\ifx \bauthor  \undefined \def \bauthor#1{#1}\fi
\ifx \batitle  \undefined \def \batitle#1{#1}\fi
\ifx \bjtitle  \undefined \def \bjtitle#1{#1}\fi
\ifx \bvolume  \undefined \def \bvolume#1{\textbf{#1}}\fi
\ifx \byear  \undefined \def \byear#1{#1}\fi
\ifx \bissue  \undefined \def \bissue#1{#1}\fi
\ifx \bfpage  \undefined \def \bfpage#1{#1}\fi
\ifx \blpage  \undefined \def \blpage #1{#1}\fi
\ifx \burl  \undefined \def \burl#1{\textsf{#1}}\fi
\ifx \doiurl  \undefined \def \doiurl#1{\url{https://doi.org/#1}}\fi
\ifx \betal  \undefined \def \betal{\textit{et al.}}\fi
\ifx \binstitute  \undefined \def \binstitute#1{#1}\fi
\ifx \binstitutionaled  \undefined \def \binstitutionaled#1{#1}\fi
\ifx \bctitle  \undefined \def \bctitle#1{#1}\fi
\ifx \beditor  \undefined \def \beditor#1{#1}\fi
\ifx \bpublisher  \undefined \def \bpublisher#1{#1}\fi
\ifx \bbtitle  \undefined \def \bbtitle#1{#1}\fi
\ifx \bedition  \undefined \def \bedition#1{#1}\fi
\ifx \bseriesno  \undefined \def \bseriesno#1{#1}\fi
\ifx \blocation  \undefined \def \blocation#1{#1}\fi
\ifx \bsertitle  \undefined \def \bsertitle#1{#1}\fi
\ifx \bsnm \undefined \def \bsnm#1{#1}\fi
\ifx \bsuffix \undefined \def \bsuffix#1{#1}\fi
\ifx \bparticle \undefined \def \bparticle#1{#1}\fi
\ifx \barticle \undefined \def \barticle#1{#1}\fi
\bibcommenthead
\ifx \bconfdate \undefined \def \bconfdate #1{#1}\fi
\ifx \botherref \undefined \def \botherref #1{#1}\fi
\ifx \url \undefined \def \url#1{\textsf{#1}}\fi
\ifx \bchapter \undefined \def \bchapter#1{#1}\fi
\ifx \bbook \undefined \def \bbook#1{#1}\fi
\ifx \bcomment \undefined \def \bcomment#1{#1}\fi
\ifx \oauthor \undefined \def \oauthor#1{#1}\fi
\ifx \citeauthoryear \undefined \def \citeauthoryear#1{#1}\fi
\ifx \endbibitem  \undefined \def \endbibitem {}\fi
\ifx \bconflocation  \undefined \def \bconflocation#1{#1}\fi
\ifx \arxivurl  \undefined \def \arxivurl#1{\textsf{#1}}\fi
\csname PreBibitemsHook\endcsname

\bibitem[\protect\citeauthoryear{Ristea et~al.}{2025}]{risteaICASSP2024Speech2025}
\begin{barticle}
\bauthor{\bsnm{Ristea}, \binits{N.-C.}},
\bauthor{\bsnm{Naderi}, \binits{B.}},
\bauthor{\bsnm{Saabas}, \binits{A.}},
\bauthor{\bsnm{Cutler}, \binits{R.}},
\bauthor{\bsnm{Braun}, \binits{S.}},
\bauthor{\bsnm{Branets}, \binits{S.}}:
\batitle{{{ICASSP}} 2024 {{Speech Signal Improvement Challenge}}}.
\bjtitle{IEEE Open J. Signal Process.}
\bvolume{6},
\bfpage{238}--\blpage{246}
(\byear{2025})
\doiurl{10.1109/OJSP.2025.3526550}
\end{barticle}
\endbibitem

\bibitem[\protect\citeauthoryear{Li et~al.}{2025}]{liASigmoidAffineProjection2025}
\begin{barticle}
\bauthor{\bsnm{Li}, \binits{Y.}},
\bauthor{\bsnm{Fu}, \binits{Y.}},
\bauthor{\bsnm{Li}, \binits{L.}},
\bauthor{\bsnm{Yu}, \binits{C.}},
\bauthor{\bsnm{Miao}, \binits{Y.}},
\bauthor{\bsnm{Zakharov}, \binits{Y.}},
\bauthor{\bsnm{Diniz}, \binits{P.S.R.}}:
\batitle{An {$\alpha$}-{{Sigmoid Affine Projection Algorithm}} for {{In-Car Echo Cancellation}} and {{Channel Estimation}}}.
\bjtitle{IEEE Trans. Veh. Technol.}
\bvolume{74}(\bissue{7}),
\bfpage{10297}--\blpage{10305}
(\byear{2025})
\doiurl{10.1109/TVT.2025.3542827}
\end{barticle}
\endbibitem

\bibitem[\protect\citeauthoryear{Huang et~al.}{2021}]{huangAffineprojectionLorentzianAlgorithm2021}
\begin{barticle}
\bauthor{\bsnm{Huang}, \binits{X.}},
\bauthor{\bsnm{Li}, \binits{Y.}},
\bauthor{\bsnm{Zakharov}, \binits{Y.V.}},
\bauthor{\bsnm{Li}, \binits{Y.}},
\bauthor{\bsnm{Chen}, \binits{B.}}:
\batitle{Affine-projection {{Lorentzian}} algorithm for vehicle hands-free echo cancellation}.
\bjtitle{IEEE Trans. Veh. Technol.}
\bvolume{70}(\bissue{3}),
\bfpage{2561}--\blpage{2575}
(\byear{2021})
\end{barticle}
\endbibitem

\bibitem[\protect\citeauthoryear{Gannot et~al.}{2017}]{gannotConsolidatedPerspectiveMultimicrophone2017}
\begin{barticle}
\bauthor{\bsnm{Gannot}, \binits{S.}},
\bauthor{\bsnm{Vincent}, \binits{E.}},
\bauthor{\bsnm{{Markovich-Golan}}, \binits{S.}},
\bauthor{\bsnm{Ozerov}, \binits{A.}}:
\batitle{A {{Consolidated Perspective}} on {{Multimicrophone Speech Enhancement}} and {{Source Separation}}}.
\bjtitle{IEEEACM Trans. Audio Speech Lang. Process.}
\bvolume{25}(\bissue{4}),
\bfpage{692}--\blpage{730}
(\byear{2017})
\doiurl{10.1109/TASLP.2016.2647702}
\end{barticle}
\endbibitem

\bibitem[\protect\citeauthoryear{Gierlich}{2014}]{gierlichCarHandsFreeTesting2014}
\begin{bchapter}
\bauthor{\bsnm{Gierlich}, \binits{H.-W.}}:
\bctitle{Car {{Hands-Free Testing}} and {{Optimization}}: {{An Overview}}}.
In: \beditor{\bsnm{Schmidt}, \binits{G.}},
\beditor{\bsnm{Abut}, \binits{H.}},
\beditor{\bsnm{Takeda}, \binits{K.}},
\beditor{\bsnm{Hansen}, \binits{J.H.L.}} (eds.)
\bbtitle{Smart {{Mobile In-Vehicle Systems}}: {{Next Generation Advancements}}},
pp. \bfpage{59}--\blpage{80}.
\bpublisher{Springer},
\blocation{New York, NY}
(\byear{2014}).
\doiurl{10.1007/978-1-4614-9120-0_5}
\end{bchapter}
\endbibitem

\bibitem[\protect\citeauthoryear{Zhang et~al.}{2024}]{zhangImprovingAcousticEcho2024}
\begin{bchapter}
\bauthor{\bsnm{Zhang}, \binits{Y.}},
\bauthor{\bsnm{Xu}, \binits{X.}},
\bauthor{\bsnm{Tu}, \binits{W.}}:
\bctitle{Improving {{Acoustic Echo Cancellation}} by {{Exploring Speech}} and {{Echo Affinity}} with {{Multi-Head Attention}}}.
In: \bbtitle{{{ICASSP}} 2024 - 2024 {{IEEE Int}}. {{Conf}}. {{Acoust}}. {{Speech Signal Process}}. {{ICASSP}}},
pp. \bfpage{401}--\blpage{405}
(\byear{2024}).
\doiurl{10.1109/ICASSP48485.2024.10446389}
\end{bchapter}
\endbibitem

\bibitem[\protect\citeauthoryear{Jiang and Biao}{2025}]{jiangSmallfootprintAcousticEcho2025}
\begin{botherref}
\oauthor{\bsnm{Jiang}, \binits{Y.}},
\oauthor{\bsnm{Biao}, \binits{T.}}:
A {{Small-footprint Acoustic Echo Cancellation Solution}} for {{Mobile Full-Duplex Speech Interactions}}.
arXiv
(2025).
\doiurl{10.48550/arXiv.2508.07561}
\end{botherref}
\endbibitem

\bibitem[\protect\citeauthoryear{Breining et~al.}{1999}]{breiningAcousticEchoControl1999}
\begin{barticle}
\bauthor{\bsnm{Breining}, \binits{C.}},
\bauthor{\bsnm{Dreiscitel}, \binits{P.}},
\bauthor{\bsnm{Hansler}, \binits{E.}},
\bauthor{\bsnm{Mader}, \binits{A.}},
\bauthor{\bsnm{Nitsch}, \binits{B.}},
\bauthor{\bsnm{Puder}, \binits{H.}},
\bauthor{\bsnm{Schertler}, \binits{T.}},
\bauthor{\bsnm{Schmidt}, \binits{G.}},
\bauthor{\bsnm{Tilp}, \binits{J.}}:
\batitle{Acoustic echo control. {{An}} application of very-high-order adaptive filters}.
\bjtitle{IEEE Signal Process. Mag.}
\bvolume{16}(\bissue{4}),
\bfpage{42}--\blpage{69}
(\byear{1999})
\doiurl{10.1109/79.774933}
\end{barticle}
\endbibitem

\bibitem[\protect\citeauthoryear{Zhang and Zhang}{2020}]{zhangRobustCascadedAcoustic2020}
\begin{bchapter}
\bauthor{\bsnm{Zhang}, \binits{C.}},
\bauthor{\bsnm{Zhang}, \binits{X.}}:
\bctitle{A {{Robust}} and {{Cascaded Acoustic Echo Cancellation Based}} on {{Deep Learning}}}.
In: \bbtitle{Interspeech 2020},
pp. \bfpage{3940}--\blpage{3944}.
\bpublisher{ISCA}, \blocation{???}
(\byear{2020}).
\doiurl{10.21437/Interspeech.2020-1260}
\end{bchapter}
\endbibitem

\bibitem[\protect\citeauthoryear{Paleologu et~al.}{2015}]{paleologuOverviewOptimizedNLMS2015a}
\begin{barticle}
\bauthor{\bsnm{Paleologu}, \binits{C.}},
\bauthor{\bsnm{Ciochin{\u a}}, \binits{S.}},
\bauthor{\bsnm{Benesty}, \binits{J.}},
\bauthor{\bsnm{Grant}, \binits{S.L.}}:
\batitle{An overview on optimized {{NLMS}} algorithms for acoustic echo cancellation}.
\bjtitle{EURASIP J. Adv. Signal Process.}
\bvolume{2015}(\bissue{1}),
\bfpage{97}
(\byear{2015})
\doiurl{10.1186/s13634-015-0283-1}
\end{barticle}
\endbibitem

\bibitem[\protect\citeauthoryear{Wang and Zhang}{2023}]{wangNewEffectiveNonparametric2023}
\begin{barticle}
\bauthor{\bsnm{Wang}, \binits{W.}},
\bauthor{\bsnm{Zhang}, \binits{H.}}:
\batitle{A new and effective nonparametric variable step-size normalized least-mean-square algorithm and its performance analysis}.
\bjtitle{Signal Process.}
\bvolume{210},
\bfpage{109060}
(\byear{2023})
\doiurl{10.1016/j.sigpro.2023.109060}
\end{barticle}
\endbibitem

\bibitem[\protect\citeauthoryear{Shamsi et~al.}{2026}]{shamsiRobustDiffusionLMS2026}
\begin{barticle}
\bauthor{\bsnm{Shamsi}, \binits{M.}},
\bauthor{\bsnm{Zayyani}, \binits{H.}},
\bauthor{\bsnm{Marvasti}, \binits{F.}}:
\batitle{Robust diffusion {{LMS}} with masked measurements}.
\bjtitle{Signal Process.}
\bvolume{238},
\bfpage{110163}
(\byear{2026})
\doiurl{10.1016/j.sigpro.2025.110163}
\end{barticle}
\endbibitem

\bibitem[\protect\citeauthoryear{Lopes}{2023}]{lopesAnalysisLMSNLMS2023}
\begin{barticle}
\bauthor{\bsnm{Lopes}, \binits{P.A.C.}}:
\batitle{Analysis of the {{LMS}} and {{NLMS}} algorithms using the misalignment norm}.
\bjtitle{Signal Image Video Process.}
\bvolume{17}(\bissue{7}),
\bfpage{3623}--\blpage{3628}
(\byear{2023})
\doiurl{10.1007/s11760-023-02588-x}
\end{barticle}
\endbibitem

\bibitem[\protect\citeauthoryear{Chen et~al.}{2024}]{chenOptimizingAcousticEcho2024}
\begin{bchapter}
\bauthor{\bsnm{Chen}, \binits{Y.-Y.}},
\bauthor{\bsnm{Wang}, \binits{J.-H.}},
\bauthor{\bsnm{Chan}, \binits{P.-C.}},
\bauthor{\bsnm{Liang}, \binits{K.-W.}},
\bauthor{\bsnm{Wang}, \binits{Z.-Y.}},
\bauthor{\bsnm{Wang}, \binits{J.-C.}}:
\bctitle{Optimizing {{Acoustic Echo Cancellation}} with {{Variable Step Size}} in {{Adaptive Filtering}}}.
In: \bbtitle{2024 {{Int}}. {{Conf}}. {{Adv}}. {{Technol}}. {{Commun}}. {{ATC}}},
pp. \bfpage{329}--\blpage{332}
(\byear{2024}).
\doiurl{10.1109/ATC63255.2024.10908122}
\end{bchapter}
\endbibitem

\bibitem[\protect\citeauthoryear{Xu et~al.}{2010}]{xuSimplifiedRLSAlgorithm2010}
\begin{bchapter}
\bauthor{\bsnm{Xu}, \binits{J.}},
\bauthor{\bsnm{Zhou}, \binits{W.-p.}},
\bauthor{\bsnm{Guo}, \binits{Y.}}:
\bctitle{A {{Simplified RLS Algorithm}} and {{Its Application}} in {{Acoustic Echo Cancellation}}}.
In: \bbtitle{2010 2nd {{Int}}. {{Conf}}. {{Inf}}. {{Eng}}. {{Comput}}. {{Sci}}.},
pp. \bfpage{1}--\blpage{4}.
\bpublisher{IEEE},
\blocation{Wuhan, China}
(\byear{2010}).
\doiurl{10.1109/ICIECS.2010.5678354}
\end{bchapter}
\endbibitem

\bibitem[\protect\citeauthoryear{Dogariu et~al.}{2022}]{dogariuIdentificationRoomAcoustic2022}
\begin{barticle}
\bauthor{\bsnm{Dogariu}, \binits{L.-M.}},
\bauthor{\bsnm{Benesty}, \binits{J.}},
\bauthor{\bsnm{Paleologu}, \binits{C.}},
\bauthor{\bsnm{Ciochin{\u a}}, \binits{S.}}:
\batitle{Identification of {{Room Acoustic Impulse Responses}} via {{Kronecker Product Decompositions}}}.
\bjtitle{IEEEACM Trans. Audio Speech Lang. Process.}
\bvolume{30},
\bfpage{2828}--\blpage{2841}
(\byear{2022})
\doiurl{10.1109/TASLP.2022.3202128}
\end{barticle}
\endbibitem

\bibitem[\protect\citeauthoryear{Otopeleanu et~al.}{2025}]{otopeleanuRobustDataReuseRegularized2025}
\begin{barticle}
\bauthor{\bsnm{Otopeleanu}, \binits{R.-A.}},
\bauthor{\bsnm{Paleologu}, \binits{C.}},
\bauthor{\bsnm{Benesty}, \binits{J.}},
\bauthor{\bsnm{Dogariu}, \binits{L.-M.}},
\bauthor{\bsnm{Stanciu}, \binits{C.-L.}},
\bauthor{\bsnm{Ciochin{\u a}}, \binits{S.}}:
\batitle{Robust {{Data-Reuse Regularized Recursive Least-Squares Algorithms}} for {{System Identification Applications}}}.
\bjtitle{Sensors}
\bvolume{25}(\bissue{16}),
\bfpage{5017}
(\byear{2025})
\doiurl{10.3390/s25165017}
\end{barticle}
\endbibitem

\bibitem[\protect\citeauthoryear{{Elisei-Iliescu} et~al.}{2019}]{elisei-iliescuRecursiveLeastSquaresAlgorithms2019}
\begin{barticle}
\bauthor{\bsnm{{Elisei-Iliescu}}, \binits{C.}},
\bauthor{\bsnm{Paleologu}, \binits{C.}},
\bauthor{\bsnm{Benesty}, \binits{J.}},
\bauthor{\bsnm{Stanciu}, \binits{C.}},
\bauthor{\bsnm{Anghel}, \binits{C.}},
\bauthor{\bsnm{Ciochin{\u a}}, \binits{S.}}:
\batitle{Recursive {{Least-Squares Algorithms}} for the {{Identification}} of {{Low-Rank Systems}}}.
\bjtitle{IEEEACM Trans. Audio Speech Lang. Process.}
\bvolume{27}(\bissue{5}),
\bfpage{903}--\blpage{918}
(\byear{2019})
\doiurl{10.1109/TASLP.2019.2903276}
\end{barticle}
\endbibitem

\bibitem[\protect\citeauthoryear{Yadav et~al.}{2025}]{yadavStateoftheartSurveyNoise2025}
\begin{barticle}
\bauthor{\bsnm{Yadav}, \binits{N.K.}},
\bauthor{\bsnm{Dhawan}, \binits{A.}},
\bauthor{\bsnm{Tiwari}, \binits{M.}},
\bauthor{\bsnm{Jha}, \binits{S.K.}}:
\batitle{A state-of-the-art survey on noise removal in a non-stationary signal using adaptive finite impulse response filtering: Challenges, techniques, and applications}.
\bjtitle{Int. J. Syst. Sci.}
\bvolume{56}(\bissue{4}),
\bfpage{885}--\blpage{918}
(\byear{2025})
\doiurl{10.1080/00207721.2024.2409850}
\end{barticle}
\endbibitem

\bibitem[\protect\citeauthoryear{Otopeleanu et~al.}{}]{otopeleanuPracticalRegularizedRecursive}
\begin{botherref}
\oauthor{\bsnm{Otopeleanu}, \binits{R.-A.}},
\oauthor{\bsnm{Benesty}, \binits{J.}},
\oauthor{\bsnm{Paleologu}, \binits{C.}},
\oauthor{\bsnm{Stanciu}, \binits{C.-L.}},
\oauthor{\bsnm{Dogariu}, \binits{L.-M.}},
\oauthor{\bsnm{Ciochina}, \binits{S.}}:
A {{Practical Regularized Recursive Least-Squares Algorithm}} for {{Robust System Identification}}
\end{botherref}
\endbibitem

\bibitem[\protect\citeauthoryear{Zakharov et~al.}{2008}]{zakharovLowComplexityRLSAlgorithms2008a}
\begin{barticle}
\bauthor{\bsnm{Zakharov}, \binits{Y.V.}},
\bauthor{\bsnm{White}, \binits{G.P.}},
\bauthor{\bsnm{Liu}, \binits{J.}}:
\batitle{Low-{{Complexity RLS Algorithms Using Dichotomous Coordinate Descent Iterations}}}.
\bjtitle{IEEE Trans. Signal Process.}
\bvolume{56}(\bissue{7}),
\bfpage{3150}--\blpage{3161}
(\byear{2008})
\doiurl{10.1109/TSP.2008.917874}
\end{barticle}
\endbibitem

\bibitem[\protect\citeauthoryear{Cioffi and Kailath}{1984}]{cioffiFastRecursiveleastsquaresTransversal1984}
\begin{barticle}
\bauthor{\bsnm{Cioffi}, \binits{J.}},
\bauthor{\bsnm{Kailath}, \binits{T.}}:
\batitle{Fast, recursive-least-squares transversal filters for adaptive filtering}.
\bjtitle{IEEE Trans. Acoust. Speech Signal Process.}
\bvolume{32}(\bissue{2}),
\bfpage{304}--\blpage{337}
(\byear{1984})
\doiurl{10.1109/TASSP.1984.1164334}
\end{barticle}
\endbibitem

\bibitem[\protect\citeauthoryear{Sutcliffe De~Moraes et~al.}{2024}]{sutcliffedemoraesFasterRLSDCDAdaptive2024}
\begin{bchapter}
\bauthor{\bsnm{Sutcliffe De~Moraes}, \binits{N.J.}},
\bauthor{\bsnm{Nascimento}, \binits{V.H.}},
\bauthor{\bsnm{Vidal}, \binits{D.C.}},
\bauthor{\bsnm{Prete}, \binits{C.A.}},
\bauthor{\bsnm{Zakharov}, \binits{Y.V.}}:
\bctitle{A faster {{RLS-DCD}} adaptive filtering algorithm}.
In: \bbtitle{2024 19th {{Int}}. {{Symp}}. {{Wirel}}. {{Commun}}. {{Syst}}. {{ISWCS}}},
pp. \bfpage{1}--\blpage{5}.
\bpublisher{IEEE},
\blocation{Rio de Janeiro, Brazil}
(\byear{2024}).
\doiurl{10.1109/ISWCS61526.2024.10639158}
\end{bchapter}
\endbibitem

\bibitem[\protect\citeauthoryear{Gouveia et~al.}{2026}]{gouveiaNumericallyStableHouseholderBased2026}
\begin{bchapter}
\bauthor{\bsnm{Gouveia}, \binits{I.M.S.}},
\bauthor{\bsnm{Apolin{\'a}rio}, \binits{J.A.}},
\bauthor{\bsnm{Saunders~F.}, \binits{C.A.B.}},
\bauthor{\bsnm{Ramos}, \binits{A.L.L.}}:
\bctitle{A {{Numerically Stable Householder-Based EX-RLS Algorithm}}}.
In: \bbtitle{{{ICASSP}} 2026 - 2026 {{IEEE Int}}. {{Conf}}. {{Acoust}}. {{Speech Signal Process}}. {{ICASSP}}},
pp. \bfpage{671}--\blpage{675}
(\byear{2026}).
\doiurl{10.1109/ICASSP55912.2026.11464615}
\end{bchapter}
\endbibitem

\bibitem[\protect\citeauthoryear{Dou et~al.}{2019}]{douFilteringTikhonovRegularizationInversion2019}
\begin{barticle}
\bauthor{\bsnm{Dou}, \binits{Z.}},
\bauthor{\bsnm{Shen}, \binits{J.}},
\bauthor{\bsnm{Li}, \binits{T.}},
\bauthor{\bsnm{Wang}, \binits{Y.}},
\bauthor{\bsnm{Gao}, \binits{M.}}:
\batitle{Filtering-{{Tikhonov}} regularization inversion for dynamic light scattering data with strong noise}.
\bjtitle{Opt. Commun.}
\bvolume{430},
\bfpage{407}--\blpage{415}
(\byear{2019})
\doiurl{10.1016/j.optcom.2018.08.078}
\end{barticle}
\endbibitem

\bibitem[\protect\citeauthoryear{Gerth}{2021}]{gerthNewInterpretationTikhonov2021}
\begin{barticle}
\bauthor{\bsnm{Gerth}, \binits{D.}}:
\batitle{A new interpretation of ({{Tikhonov}}) regularization}.
\bjtitle{Inverse Probl.}
\bvolume{37}(\bissue{6}),
\bfpage{064002}
(\byear{2021})
\doiurl{10.1088/1361-6420/abfb4d}
\end{barticle}
\endbibitem

\bibitem[\protect\citeauthoryear{Saremi et~al.}{2023}]{saremiAcousticEchoCanceller2023}
\begin{barticle}
\bauthor{\bsnm{Saremi}, \binits{A.}},
\bauthor{\bsnm{Ramkumar}, \binits{B.}},
\bauthor{\bsnm{Ghaffari}, \binits{G.}},
\bauthor{\bsnm{Gu}, \binits{Z.}}:
\batitle{An acoustic echo canceller optimized for hands-free speech telecommunication in large vehicle cabins}.
\bjtitle{EURASIP J. Audio Speech Music Process.}
\bvolume{2023}(\bissue{1}),
\bfpage{39}
(\byear{2023})
\doiurl{10.1186/s13636-023-00305-7}
\end{barticle}
\endbibitem

\bibitem[\protect\citeauthoryear{Tylavsky and Sohie}{1986}]{tylavskyGeneralizationMatrixInversion1986}
\begin{barticle}
\bauthor{\bsnm{Tylavsky}, \binits{D.J.}},
\bauthor{\bsnm{Sohie}, \binits{G.R.L.}}:
\batitle{Generalization of the matrix inversion lemma}.
\bjtitle{Proc. IEEE}
\bvolume{74}(\bissue{7}),
\bfpage{1050}--\blpage{1052}
(\byear{1986})
\doiurl{10.1109/PROC.1986.13587}
\end{barticle}
\endbibitem

\bibitem[\protect\citeauthoryear{Clark et~al.}{1981}]{clarkBlockImplementationAdaptive1981}
\begin{barticle}
\bauthor{\bsnm{Clark}, \binits{G.}},
\bauthor{\bsnm{Mitra}, \binits{S.}},
\bauthor{\bsnm{Parker}, \binits{S.}}:
\batitle{Block implementation of adaptive digital filters}.
\bjtitle{IEEE Trans. Acoust. Speech Signal Process.}
\bvolume{29}(\bissue{3}),
\bfpage{744}--\blpage{752}
(\byear{1981})
\end{barticle}
\endbibitem

\bibitem[\protect\citeauthoryear{Fayadh et~al.}{2014}]{fayadhEnhancementThreeCombining2014}
\begin{barticle}
\bauthor{\bsnm{Fayadh}, \binits{R.A.}},
\bauthor{\bsnm{Malek}, \binits{F.}},
\bauthor{\bsnm{Fadhil}, \binits{H.A.}}:
\batitle{Enhancement of a {{Three Combining Techniques Rake Receiver Using Adaptive Filter}} of {{M-Max Partial Update RLS Algorithm}} for {{DS-UWB Systems}}}.
\bjtitle{J. Gener. Inf. Technol.}
\bvolume{5}(\bissue{4}),
\bfpage{93}
(\byear{2014})
\end{barticle}
\endbibitem

\bibitem[\protect\citeauthoryear{Sridhar et~al.}{2021}]{sridharICASSP2021Acoustic2021}
\begin{bchapter}
\bauthor{\bsnm{Sridhar}, \binits{K.}},
\bauthor{\bsnm{Cutler}, \binits{R.}},
\bauthor{\bsnm{Saabas}, \binits{A.}},
\bauthor{\bsnm{Parnamaa}, \binits{T.}},
\bauthor{\bsnm{Loide}, \binits{M.}},
\bauthor{\bsnm{Gamper}, \binits{H.}},
\bauthor{\bsnm{Braun}, \binits{S.}},
\bauthor{\bsnm{Aichner}, \binits{R.}},
\bauthor{\bsnm{Srinivasan}, \binits{S.}}:
\bctitle{{{ICASSP}} 2021 {{Acoustic Echo Cancellation Challenge}}: {{Datasets}}, {{Testing Framework}}, and {{Results}}}.
In: \bbtitle{{{ICASSP}} 2021 - 2021 {{IEEE Int}}. {{Conf}}. {{Acoust}}. {{Speech Signal Process}}. {{ICASSP}}},
pp. \bfpage{151}--\blpage{155}
(\byear{2021}).
\doiurl{10.1109/ICASSP39728.2021.9413457}
\end{bchapter}
\endbibitem

\bibitem[\protect\citeauthoryear{Dogariu et~al.}{2022}]{dogariuPerformanceDataReuseFast2022}
\begin{bchapter}
\bauthor{\bsnm{Dogariu}, \binits{L.-M.}},
\bauthor{\bsnm{Paleologu}, \binits{C.}},
\bauthor{\bsnm{Benesty}, \binits{J.}},
\bauthor{\bsnm{Ciochin{\u a}}, \binits{S.}}:
\bctitle{On the {{Performance}} of a {{Data-Reuse Fast RLS Algorithm}} for {{Acoustic Echo Cancellation}}}.
In: \bbtitle{2022 {{IEEE Int}}. {{Black Sea Conf}}. {{Commun}}. {{Netw}}. {{BlackSeaCom}}},
pp. \bfpage{135}--\blpage{140}
(\byear{2022}).
\doiurl{10.1109/BlackSeaCom54372.2022.9858200}
\end{bchapter}
\endbibitem

\end{thebibliography}

\end{document}